\documentclass[preprint,tighten,floats,twoside,eqsecnum,aps,epsf,epsfig]{revtex4}
\usepackage{graphicx}

\usepackage{color}
\usepackage{bm}
\newcommand{\beq}{\begin{equation}}
\newcommand{\enq}{\end{equation}}
\newcommand{\beqa}{\begin{eqnarray}}
\newcommand{\beqast}{\begin{eqnarray*}}
\newcommand{\enqa}{\end{eqnarray}}
\newcommand{\enqast}{\end{eqnarray*}}

\begin{document}

\title{ Amplitudes and Observables in pp Elastic Scattering \\
at $\sqrt{s}=7$ TeV }
 
  \author{A. K. Kohara} 
 \affiliation{Instituto de F\'{\i}sica, Universidade Federal do Rio de
 Janeiro \\
 C.P. 68528, Rio de Janeiro 21945-970, RJ, Brazil   }
\author{E. Ferreira} 
 \affiliation{Instituto de F\'{\i}sica, Universidade Federal do Rio de
 Janeiro \\
 C.P. 68528, Rio de Janeiro 21945-970, RJ, Brazil   }
\author{T. Kodama} 

\altaffiliation{EMMI at FIAS-Frankfurt Institute for Advanced Study, Ruth-Moufang Str. 1, 60438, Frankfurt am Main, Germany} 
 \affiliation{Instituto de F\'{\i}sica, Universidade Federal do Rio de
 Janeiro \\
 C.P. 68528, Rio de Janeiro 21945-970, RJ, Brazil   }


\begin{abstract}

A precise analysis of the pp elastic scattering data at 7 TeV in terms of its
amplitudes is performed as  an  extension of previous studies for lower
energies. Slopes $B_{R}$ and $B_{I}$ of the real and imaginary amplitudes are 
independent quantities, and  a proper expression for the Coulomb phase is used. 
The real and imaginary amplitudes are fully disentangled,
consistently with forward dispersion relations for amplitudes and for slopes.
We present analytic expressions for the amplitudes that cover all $t$ range 
completely, while values of total cross section $\sigma$, ratio $\rho$, 
$B_{I}$ and $B_{R}$ enter consistently to describe forward scattering. 
 It is stressed that the identification of the amplitudes is an essential step
for the description of elastic scattering, and pointed out the importance of
the experimental investigation of the transition range from non-perturbative
to perturbative dynamics, that may confirm the three gluon exchange mechanism 
observed at lower energies.


\end{abstract}
\maketitle





\section{Introduction \label{intro}}

The determination of the imaginary and real parts of the amplitudes of elastic
pp and $\mathrm{{p\bar{p}}}$ scattering is an essential step for the
understanding of the underlying dynamics governing hadronic interactions and
for the construction and critical evaluation of theoretical models. This
question has now increased renewed interest in view of the experimental
activity at new energy frontiers of the LHC, particularly with the forward physics
measurements of the TOTEM Collaboration \cite{TOTEM, talks}, and also with the
possibility of measurements in the large $|t|$ range. However, the
identification of amplitudes is not at all a trivial task. There is no 
methodology  completely free from  subjective judgment and of interpolations 
and extrapolations based on investigation of data at all energies.

Efforts to identify amplitudes were made in a treatment originated in the
Stochastic Vacuum Model \cite{dosch}, resulting in good  description of
the data on $d\sigma/dt$ at all energies from 20 to 1800 GeV \cite{ferreira1}.
In particular, it was made clear to establish the regularity in the
general features of the real and imaginary amplitudes, with identification 
of magnitudes, signs and zeros of the individual parts.

The constraints imposed by the well-known dispersion relations for the
amplitudes and also by the new dispersion relations for the slopes
\cite{ferreira2} help to control the forward scattering parameters, and are
an essential tool for the construction of valid analytic representations of the
data. In the present work, we extend the previous analysis to the recent pp
elastic scattering measurements of the LHC TOTEM Collaboration \cite{TOTEM,
talks} at $\sqrt{s}=7\ $ TeV, obtaining  a very precise description of the 
amplitudes and of the differential cross section in the whole $|t|$ range,
including location, depth and shape of dip and bump. The quantities $\sigma$ ,
$\rho$ and slopes $B_{I}$ and $B_{R} $ that are the common parameters of the
basic features of the forward scattering data, are also determined precisely
from this all-$t$ analysis. Furthermore, we show that the 7 TeV data favor the
assumption of universality of the contribution from the perturbative three
gluon exchange process at large $|t| $ that has been observed at 27.4 GeV
\cite{Faissler} and confirmed at higher ISR energies.

Our analysis shows why the dip/bump or inflection structure, that appears in pp
and p\={p} elastic differential cross section at ISR and Tevatron energies,  has
also been observed at 7 TeV \cite{TOTEM, talks}. This is due to the monotonic
displacement of the zero of the imaginary part as the incident energy
increases, approaching the first zero of the real amplitude, with formation of
a marked dip at about 0.5 GeV$^2$. We remark that at the energies 540 and 1800 GeV 
no dip-bump appears  because the imaginary zero is in the middle
between the two zeros of the real part, with compensating contributions. This
analysis also suggests that, mainly as consequence of the continued
displacement of the zero of the imaginary part, an even more pronounced
dip/bump structure should appear for $14$ TeV pp scattering at a slightly
smaller value $|t|\simeq0.4\nobreak\,\mbox{GeV}^{2} $.

We organize the present paper as follows. In Sec. \ref{amplitudes_forms}, we
introduce the representation of our amplitudes. This representation contains 
the exponential nature of the very forward domain, with associated Coulomb amplitude and
phase, and the larger $|t|$ behavior with dip and bump, together with the
contribution from the perturbative three-gluon exchange process, all in unique
analytic forms. We also show the form of the nuclear amplitude in impact
parameter space, originated in studies with the Stochastic Vacuum Model. In
Sec. \ref{data_7TeV} we show our results for the scattering amplitudes
obtained from a precise fit to the experimental data at 7 TeV. It is shown that the
forward scattering parameters determined from our analytic expression for all
$t$ values are in agreement with those determined by the TOTEM collaboration
\cite{talks} . In Sec. \ref{results_amplitudes} we discuss the physical nature
of the amplitudes, the position of zeros and its relation to the observed
dip/bump structure. These properties are compared to the 52.8 GeV case, where a wider $t$ domain has been measured. To make clear the physical meaning of the properties of the amplitudes, a comparison with another model is made.
Finally in Sec. \ref{summary} we
give a summary of the present work, and discuss the consequences of our
analysis, together with the future perspectives. In Appendix, we derive the
Coulomb phase function extended to deal with the case of different 
$B_{I}$\ and $B_{R}$ slopes . \bigskip


\section{Analytic Representation of Amplitudes of $\mathrm{pp}$ elastic
scattering \label{amplitudes_forms}}

We write the differential cross section as
\begin{equation}
\frac{d\sigma}{dt}= \left(  \hbar c\right)  ^{2}~|T_{R}(s,t)+iT_{I}%
(s,t)|^{2}~.
\end{equation}
In the following discussion, we use the unit system where $\sigma$ is in mb
(milibarns) and the energy in GeV, so that $\left(  \hbar c\right)
^{2}=0.3894\ $mb~GeV$^{2}$.

$T_{R}$ and $T_{I}$, with dimensions GeV$^{-1}$, contain the nuclear and the
Coulomb parts in the forms
\begin{equation}
T_{R}(s,t)= T_{R}^{N}(s,t) +\sqrt{\pi}F^{C}(t)\cos(\alpha\Phi)~,\label{real}%
\end{equation}
and
\begin{equation}
T_{I}(s,t)= T_{I}^{N}(s,t) +\sqrt{\pi}F^{C}(t)\sin(\alpha\Phi)~.\label{imag}%
\end{equation}
The Coulomb scattering amplitude $F^{C}(s,t)$ enters in the pp$/$%
p$\mathrm{{\bar{p}}}$ amplitudes with the form
\begin{equation}
F^{C}(t)~ e^{i\alpha\Phi(s,t)} =(-/+)~\frac{2\alpha}{|t|}~F_{\mathrm{proton}%
}^{2}(t) ~ e^{i\alpha\Phi(s,t)} ~,\label{coulomb}%
\end{equation}
where $\alpha~$is the fine-structure constant, $\Phi(s,t)$ is the Coulomb
phase and the proton form factor is written
\begin{equation}
F_{\mathrm{proton}}(t)=[0.71/(0.71+|t|)]^{2}~.\label{ff_proton}%
\end{equation}
The total cross section is given by
\begin{equation}
\sigma= 4 \sqrt{\pi} \left(  \hbar c\right) ^{2}~ T_{I}^{N}(s,t=0) ~.
\end{equation}

$T_{R}^{N}(s,t)$ and $T_{I}^{N}(s,t)$, respectively the real and imaginary
parts of the properly normalized scattering amplitude of the strong
interaction, are smooth and regular functions of $s$ and $t$, and the Coulomb
amplitude is relevant in the very forward range $|t| < 10^{-2}~ \mathrm{{GeV}%
^{2} }$. We neglect spin effects.

\subsection{ Nuclear Amplitudes for all-$t$ Values}

To obtain precise description of the elastic $d\sigma/dt$ data for all $|t|$,
we use the forms that have been introduced before and 
shown to be successful with ISR and
Fermilab data \cite{ferreira1}, together with the assumption of the
perturbative three-gluon exchange amplitude \cite{landshoff}, writing
\begin{equation}
T_{K}^{N}(s,t)=\alpha_{K}(s)\mathrm{e}^{-\beta_{K}(s)|t|}+\lambda_{K}%
(s)\Psi_{K}(\gamma_{K}(s),t)+\delta_{K,R}R_{ggg}\left(  t\right)
~,\label{ampTK}%
\end{equation}
where $\Psi_{K}(\gamma_{K}(s),t)$ are shape functions, described below, and
$R_{ggg}\left(  t\right)  $ represents the contribution from the perturbative
three-gluon exchange amplitude. The label $K$ means either $K=R $ for the real
amplitude or $K=I$ for the imaginary amplitude, and $\delta_{K,R}$ is the
Kronecker%
\'{}%
s delta symbol, that is, the last term only contributes for the real part.

\subsubsection{Shape functions}

The shape functions are written
\begin{equation}
\Psi_{K}(\gamma_{K}(s),t)=2~\mathrm{e}^{\gamma_{K}}~\bigg[{\frac
{\mathrm{e}^{-\gamma_{K}\sqrt{1+a_{0}|t|}}}{\sqrt{1+a_{0}|t|}}}-\mathrm{e}%
^{\gamma_{K}}~{\frac{e^{-\gamma_{K}\sqrt{4+a_{0}|t|}}}{\sqrt{4+a_{0}|t|}}%
}\bigg]~,\label{cd1_s}%
\end{equation}
with the property $\Psi_{K}(\gamma_{K}(s),t=0) = 1 $. We have here introduced 
for each amplitude 4 energy dependent parameters, $\alpha_{K}$, $\beta_{K}$,
$\gamma_{K}$, $\lambda_{K}$,
whose roles are explained below. $\gamma_{K}$ is dimensionless, while
$\alpha_{K}$, $\gamma_{K}$ and $\beta_{K}$ are in GeV$^{-2}$. The fixed
quantity $a_{0}=1.39$ GeV$^{-2}$ is related to the square of the correlation
length $a$ of the gluon vacuum expectation value, with $~a~=(0.2\sim0.3)$ fm ~
\cite{dosch}.

These forms for the amplitudes originate in a formulation of pp scattering in
impact parameter $b$ space in studies based on the Stochastic Vacuum Model
\cite{dosch,ferreira1} . The analytic representation of our nuclear amplitude
(except for the perturbative three-gluon exchange contribution) has a simple
form in terms of impact parameter space through the Fourier transforms,
\[
\tilde{T}_{K}(s,b)=\frac{1}{2\pi}\int d^{2}\vec{q}~e^{-i\vec{q}.\vec{b}}%
~T_{K}^{N}(s,t=-q^{2}),~
\]
which are given in closed forms as
\begin{equation}
\tilde{T}_{K}(s,b)=\frac{\alpha_{K}}{2~\beta_{K}}e^{-\frac{b^{2}}{4\beta_{K}}%
}+\lambda_{K}~\tilde{\psi}_{K}(s,b)~,\label{Eikonal}%
\end{equation}
where
\begin{equation}
\tilde{\psi}_{K}(s,b)=\frac{2~e^{\gamma_{K}}}{a_{0}}~\frac{e^{-\sqrt
{\gamma_{K}^{2}+\frac{b^{2}}{a_{0}}}}}{\sqrt{\gamma_{K}^{2}+\frac{b^{2}}%
{a_{0}}}}\Big[1-e^{\gamma_{K}}~e^{-\sqrt{\gamma_{K}^{2}+\frac{b^{2}}{a_{0}}}%
}\Big]~\label{Eikonal2}%
\end{equation}
are the shape functions in $b$ space. More details of the representation of
the amplitudes in $b$-space are given in Sec. \ref{results_amplitudes}. We here
choose to work with the amplitudes in $t$-space because they are more directly
connected to the $d\sigma/dt$ data that we intend to describe.

These expressions are planned to represent the non - perturbative dynamics  of 
scattering for all $|t|$  and the perturbative term $R_{ggg}$ is tuned 
to vanish for small $|t|$.

The limits at $|t|$ = 0 lead to the values for
the total cross section $\sigma$, the ratio $\rho$ of the real to imaginary
amplitudes, and the slopes $B_{R,I}$ at $t=0$ through
\begin{equation}
\sigma(s)=4\sqrt{\pi}\left(  \hbar c\right)  ^{2}~(\alpha_{I}(s)+\lambda
_{I}(s))~,\label{sigma_par}%
\end{equation}%
\begin{equation}
\rho(s)=\frac{T_{R}^{N}(s,t=0)}{T_{I}^{N}(s,t=0)}=\frac{\alpha_{R}%
(s)+\lambda_{R}(s)}{\alpha_{I}(s)+\lambda_{I}(s)}~,\label{rho_par}%
\end{equation}
and 
\begin{eqnarray}
B_{K}(s) =\frac{1}{T_{K}^{N}(s,t)}\frac{dT_{K}^{N}(s,t)}{dt}\Big|_{t=0}%
=~\frac{1}{\alpha_{K}(s)+\lambda_{K}(s)}\times\nonumber\\
\Big[\alpha_{K}(s)\beta_{K}(s)+\frac{1}{8}\lambda_{K}(s)a_{0}\Big(6\gamma
_{K}(s)+7\Big)\Big]~ .\label{slopes_par}%
\end{eqnarray}

The determination of each amplitude starts with four energy dependent
parameters that must be obtained from the data. However, with well established
$s$ dependences of the imaginary slope $B_{I}$ (with typical Regge model
behavior), connections between $\lambda_{R}+\alpha_{R}$ and $\lambda
_{I}+\alpha_{I}$, plus constraints from dispersion relation for slopes
\cite{ferreira2}, in practice we deal with fewer independent parameters to
describe $d\sigma/dt$ at each energy.

As discussed before \cite{ferreira1}, the shape functions in Eq. (\ref{cd1_s}) 
are very convenient choices for all
values of $|t|$, determining consistently for all energies the zeros, the
formation of dips and bumps, the signs and magnitudes of the two amplitudes
and are able to reproduce with good accuracy all $d\sigma/dt$ behavior. This
description represents the non-perturbative QCD dynamics that is responsible
for soft elastic hadronic scattering. They account effectively for the terms
of Regge and/or eikonal phenomenology that determine the process for $|t|$
ranges up to about $|t|\approx2.5$ GeV$^{2}$. For higher $|t|$, perturbative
contributions may become important.

\subsubsection{Universal behavior at large $|t|$ : Faissler measurements at
27.4 GeV}

For $|t|$ values beyond the dip and bump characteristic of the differential
cross sections at ISR/CERN and Fermilab energies, namely for $|t| > 1.5 ~
\mathrm{{GeV}^{2}}$, the differential cross sections become increasingly
independent of the energy. The measurements of pp scattering at $\sqrt{s}=$
27.4 GeV \cite{Faissler} provides the only large $|t|$ data, covering the
interval from 5.5 to 14.2 GeV$^{2}$. It is known that the points are smoothly
and naturally connected with the lower $|t|$ points at all energies
\cite{ferreira1}. As shown by Donnachie and Landshoff \cite{landshoff} this
tail corresponds to a perturbative three-gluon exchange mechanism, with real
amplitude contribution positive for pp and negative for $\mathrm{{p\bar p} }$ scattering,
and $d\sigma/dt$ falling as $|t|^{-8}$. The most remarkable example is given
by the data of pp scattering at $\sqrt{s}$ = 52.8 GeV \cite{data_52} , with
measurements up to $|t|=9.75~\mathrm{{GeV}^{2}}$ that superpose well with the
27.4 GeV tail. The observed behavior indicates that in this region the non-perturbative real
part is indeed positive, and also that the magnitude of the imaginary part is
small compared to the real part. At high energies in $\mathrm{p \bar p}$
scattering the negative sign of the perturbative tail may lead to a marked dip
in the transition region from 3 to 4 GeV$^{2}$.

Based on this expectation for the perturbative tail, we introduce a simple
form of the universal three-gluon contribution amplitude $R_{ggg}(t)$ at large
$|t|$, parameterizing the Landshoff term of the amplitude \cite{landshoff} as
\begin{equation}
R_{ggg}(t)\equiv\pm0.45~t^{-4}(1-e^{-0.005\left\vert t\right\vert ^{4}%
})(1-e^{-0.1\left\vert t\right\vert ^{2}})~,\label{R_tail}%
\end{equation}
where the last two factors cut-off  this term smoothly in the non-perturbative
domain, and the signs $\pm$ refer to the pp and p$\mathrm{{\bar{p}}}$
amplitudes respectively.

Although the cut-off factors written in Eq. (\ref{R_tail}) have been adequate 
for all cases that were examined, their detailed forms in the transition range 
$(2.5~<~|t|~<~4)$ ~ GeV$^{2}$  must be examined with data.

\subsection{Exponential forms of amplitudes for very small $|t|$}

In the treatment of elastic pp and p$\mathrm{{\bar{p}}}$ scattering in the
forward direction, with amplitudes approximated by pure exponential forms, the
differential cross section is written
\begin{eqnarray}
\frac{d\sigma}{dt}  & =\pi\left(  \hbar c\right)  ^{2}~~\Big\{\Big[\frac
{\rho\sigma}{4\pi\left(  \hbar c\right)  ^{2}}~{{e}^{B_{R}t/2}+F^{C}%
(t)\cos{(\alpha\Phi)}\Big]^{2}}\nonumber\\
& +\Big[\frac{\sigma}{4\pi\left(  \hbar c\right)  ^{2}}~{{e}^{B_{I}t/2}%
+F^{C}(t)\sin{(\alpha\Phi)}\Big]^{2}\Big\}~,}\label{diffcross_eq}%
\end{eqnarray}
where $t\equiv-|t|$ and we account for different values for the slopes $B_{I}$
and $B_{R}$ of the imaginary and real amplitudes.  
The expression for the Coulomb phase that accounts for 
  $B_{R}\neq B_{I}$ derived in the Appendix.

In elastic pp and p$\mathrm{\bar p}$ scattering at all energies above $\sqrt
s=$ 19 GeV, the real and imaginary amplitudes have zeros located in ranges
$|t|\approx(0.1 \sim0.3)$ GeV$^{2}$ and $|t|=(0.5 \sim1.5)$ GeV$^{2}$
respectively, and the use of exponential forms beyond a limited forward range
could lead to inaccurate determination of the characteristic forward scattering
parameters $\sigma$, $\rho$, $B_{I}$ and $B_{R}$. We then need to use more
general expressions that are connected with the exponential behavior as
limits, such as the forms with shape functions written in Eqs. (\ref{ampTK})
and (\ref{cd1_s}).

\section{ Data on $\mathrm{pp}$ scattering at 7 TeV \label{data_7TeV}}

\subsection{Determination of Amplitude Parameters}

\begin{figure}[ptb]
\caption{ Analytic representation for the data on $d\sigma/dt$ for elastic pp
scattering at 7 TeV, together with the experimental data \cite{talks}; 
triangles are for subset $A$ and open squares for subset $B$. 
The values of the parameters are given 
in Eqs. (\ref{input}), (\ref{other_param}). The inset at the up-right 
corner shows in closeup a gap between the datasets $A$  and  $B$, that seems 
to be at the  limit of the error bars.  } 
    \includegraphics[height=7.0cm]{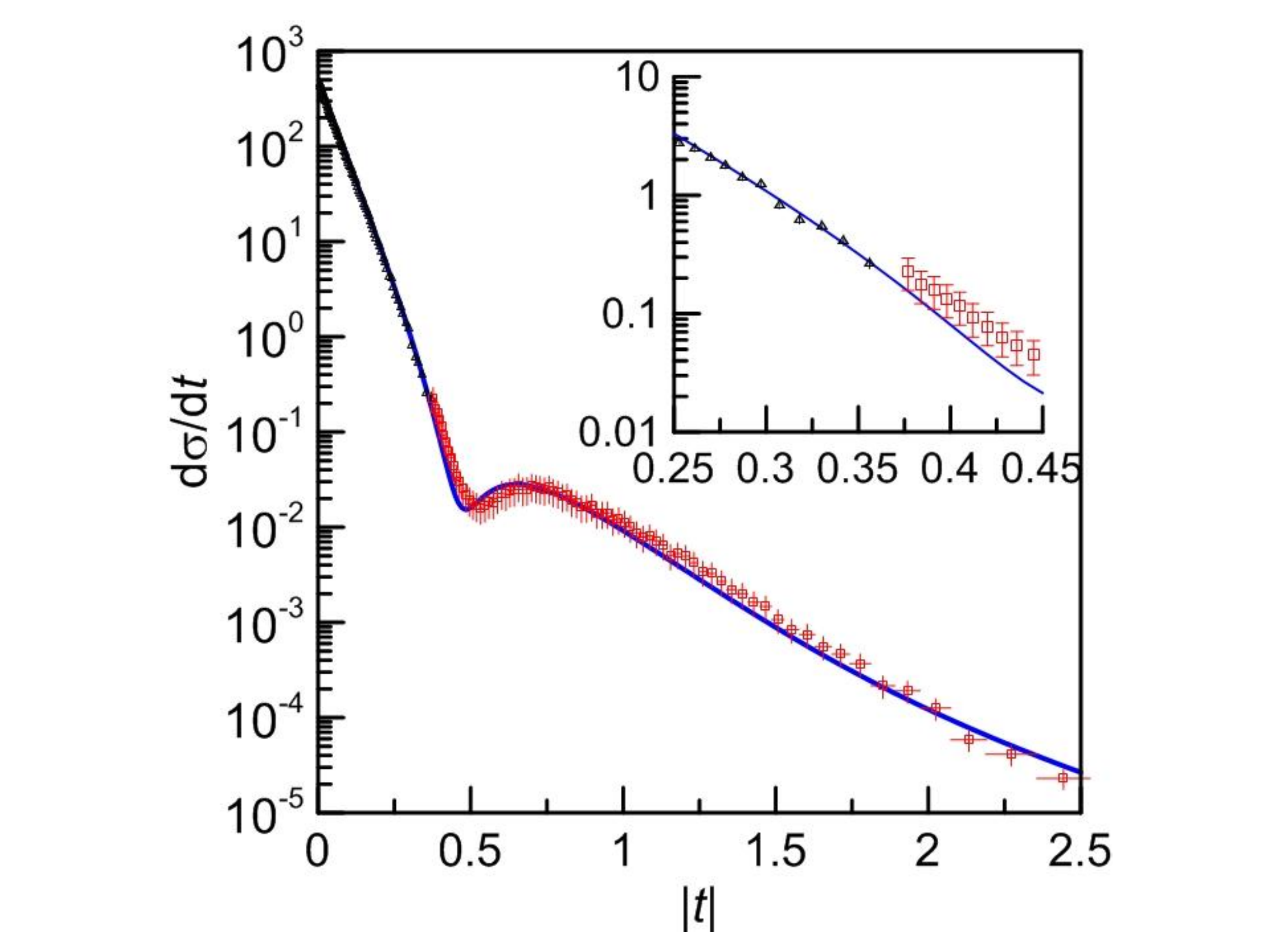}
\label{cross_0_7TeV}%
\end{figure}

The TOTEM Collaboration has published data of differential elastic cross
section \cite{talks} in two separate tables, in the ranges $0.00515\leq
|t|\leq0.371$ (referred to as dataset $A$ )
and $0.377\leq|t|\leq2.443$ (referred to as dataset $B$ )  
 $\nobreak\,\mbox{GeV}^{2}$, with 87
and 78 points respectively. Rather large systematic errors are informed in the dataset $B$. The representation of these data with the analytical
forms of Eqs. (\ref{ampTK}) and (\ref{cd1_s}) is shown in Fig.
\ref{cross_0_7TeV}. The averaged square deviation for the dataset $A+B$ (165 points) is
$\langle\chi^{2}\rangle=0.3105$. 
\begin{figure}[ptb]
 \caption{ Data for elastic pp scattering at 7 TeV (\cite{talks}, 165 points)
and 52.8 GeV ( \cite{data_52}, 97 points). The lines represent the
parameterizations with Eqs. (\ref{ampTK}) and (\ref{cd1_s}), with average
squared deviations $\langle\chi^{2}\rangle$ = 0.3105 and 0.8328 respectively.
Numerical characteristic values and description of features of the data and of
the amplitudes are given in the text and in Tables \ref{tableone} and
{\protect {\ref{tabletwo}} } . }%
\label{plot_52_7TeV}
\includegraphics[height=10.2cm]{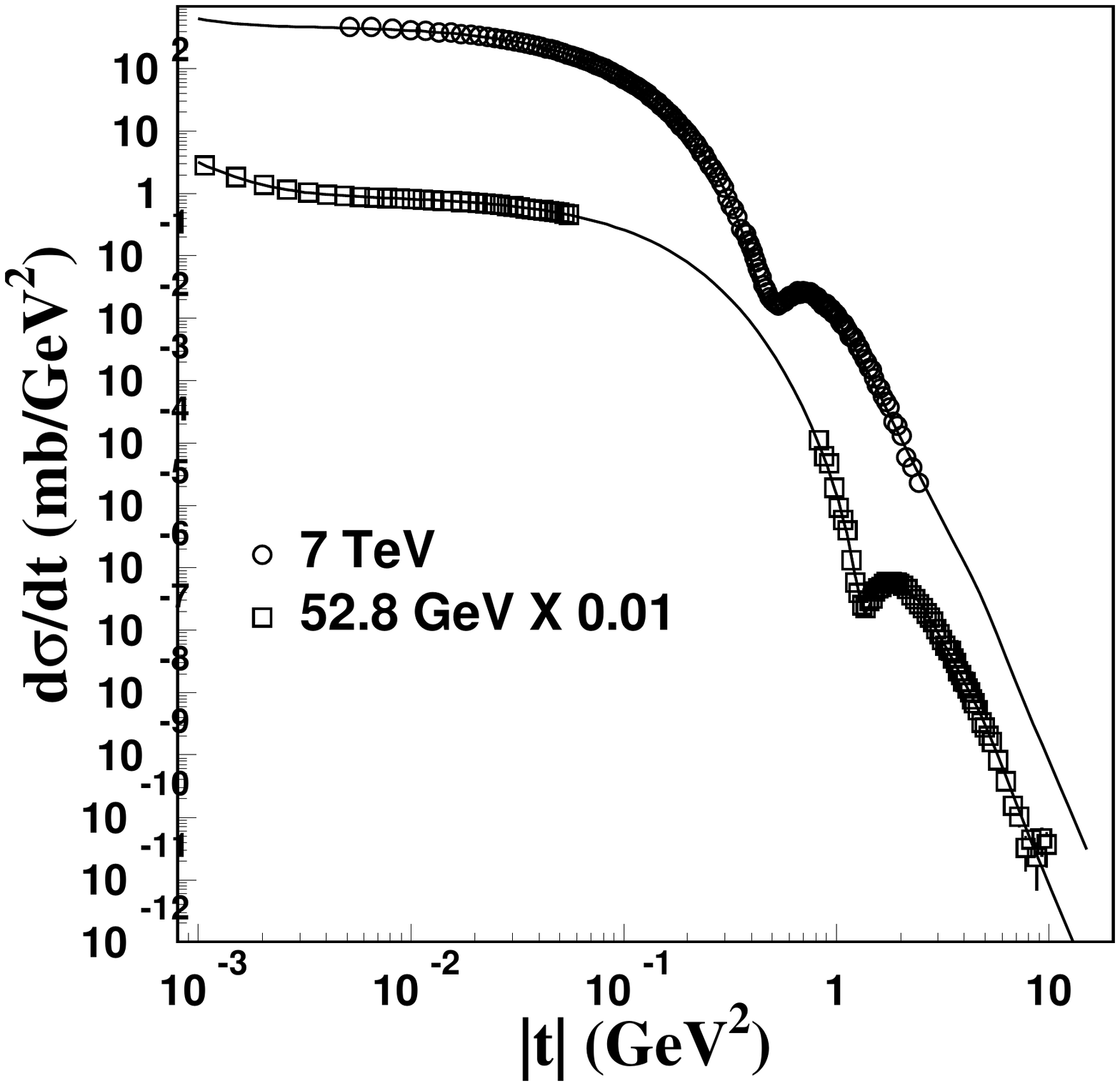}
\end{figure}
The values of the quantities that also enter in the forward scattering
amplitudes treated with exponential functions as in Eq. (\ref{diffcross_eq}) are
\begin{eqnarray}
& \sigma=98.65\pm0.26~\mathrm{mb}~,~\rho=0.141\pm0.001~,\nonumber\\
& B_{I}=19.77\pm0.03~\mathrm{GeV}^{-2},~B_{R}=30.2\pm0.7~\mathrm{GeV}%
^{-2}~, \label{input}%
\end{eqnarray}
and the other independent quantities that enter in the full-t forms are (all in
GeV$^{-2}$ )
\begin{eqnarray}
&  \alpha_{I}=13.730\pm0.030~,~\beta_{I}=4.0826\pm0.0093~,\nonumber\\
&  \lambda_{R}=4.7525\pm0.0155~,~\beta_{R}=1.4851\pm
0.0318~.\label{other_param}%
\end{eqnarray}
Here we have 8 independent values, that are collected in Table \ref{tableone}.
\footnote{Because the present data stops at about 2.5
$\nobreak\,\mbox{GeV}^{2}$, the quantity $\beta_{R}$ is difficult to fix
uniquely.} Other quantities are related to these through Eqs. (\ref{sigma_par}%
, \ref{rho_par}, \ref{slopes_par}). Plots and features of the amplitudes are
given in the next section. 
\begin{table*}[ptb]
\caption{Amplitude parameters for the 52.8 GeV and 7 TeV data.    }
  \label{tableone}
\begin{tabular*}{\textwidth}{@{\extracolsep{\fill}}lcccccccccc@{}}
\hline
$\sqrt s$& N  &$\sigma $ ~ & $\rho$~ & $B_I$~  & $B_R$ & $\alpha_I$  &$\beta_I$&$\lambda_R$ & $\beta_R$ & $\langle\chi^2\rangle$\\ 
GeV & points &  mb  & -- & GeV$^{-2}$ & GeV$^{-2}$ & GeV$^{-2}$ & GeV$^{-2}$  & GeV$^{-2}$  & GeV$^{-2}$ &    \\ \hline
 52.8& 97 & 42.49 & 0.078    & 13.04 & 19.07 & 5.9561 & 2.3477 &1.1307   & 1.1436 & 0.8328 \\
7000& 165   & 98.65  &0.141  &19.77  &30.20 & 13.730   &4.0826 & 4.7525 & 1.4851  & 0.3105 \\ \hline 
\end{tabular*}
\end{table*}

Characteristic values of the differential and integrated cross sections for
the 7 TeV data given by this representation are
\begin{eqnarray}
&  \sigma_{\mathrm{el.}} = 25.5418 ~ \mathrm{mb } ~ ; ~ \sigma
_{\mathrm{\ inel.}} = 73.1082 ~ \mathrm{mb } ~ ; ~ \sigma_{\mathrm{el.}%
}/\sigma= 0.2589 ~ ;\nonumber\\
& |t|_{\mathrm{dip}}= 0.4847 ~ \nobreak\,\mbox{GeV}^{2}~ ;~ (d\sigma
/dt)_{\mathrm{dip}}= 0.01532 ~ \mathrm{{mb} }~\nobreak\,\mbox{GeV}^{-2}
;\nonumber\\
& |t|_{\mathrm{bump}} = 0.6488 ~ \nobreak\,\mbox{GeV}^{2} ; (d\sigma
/dt)_{\mathrm{bump}} = 0.02816 ~ \mathrm{mb}~\nobreak\,\mbox{GeV}^{-2} ~
;\nonumber\\
& \mathrm{ratio (bump/dip)} = 1.8383 .~~~~~~~~~~\label{features}%
\end{eqnarray}

In a closeup examination of the graph, shown in the inset in 
Fig. \ref{cross_0_7TeV}, we observe a discontinuity in the magnitudes of
$d\sigma/dt$  in the junction of the two sets of data. We
estimate that a renormalization factor 0.86  would adjust the higher $|t|$ 
to the more forward data in this region. 

 It is very interesting to compare the 7 TeV data with the similar behavior of
the 52.8 GeV data \cite{data_52}, that are available in the ranges
$0.107\times10^{-2} \leq|t| \leq0.5546 \times10^{-2} ~
\nobreak\,\mbox{GeV}^{2} $ with 34 points and $0.825 \leq|t| \leq9.75 ~
\nobreak\,\mbox{GeV}^{2} $ with 63 points. This comparison is shown in Fig.
\ref{plot_52_7TeV} and in Tables \ref{tableone} and \ref{tabletwo}, and the
behavior of the amplitudes is discussed in Sec. \ref{results_amplitudes}.

\section{ Amplitudes \label{results_amplitudes}}

\begin{figure*}[ptb]
\includegraphics[height=5.8cm]{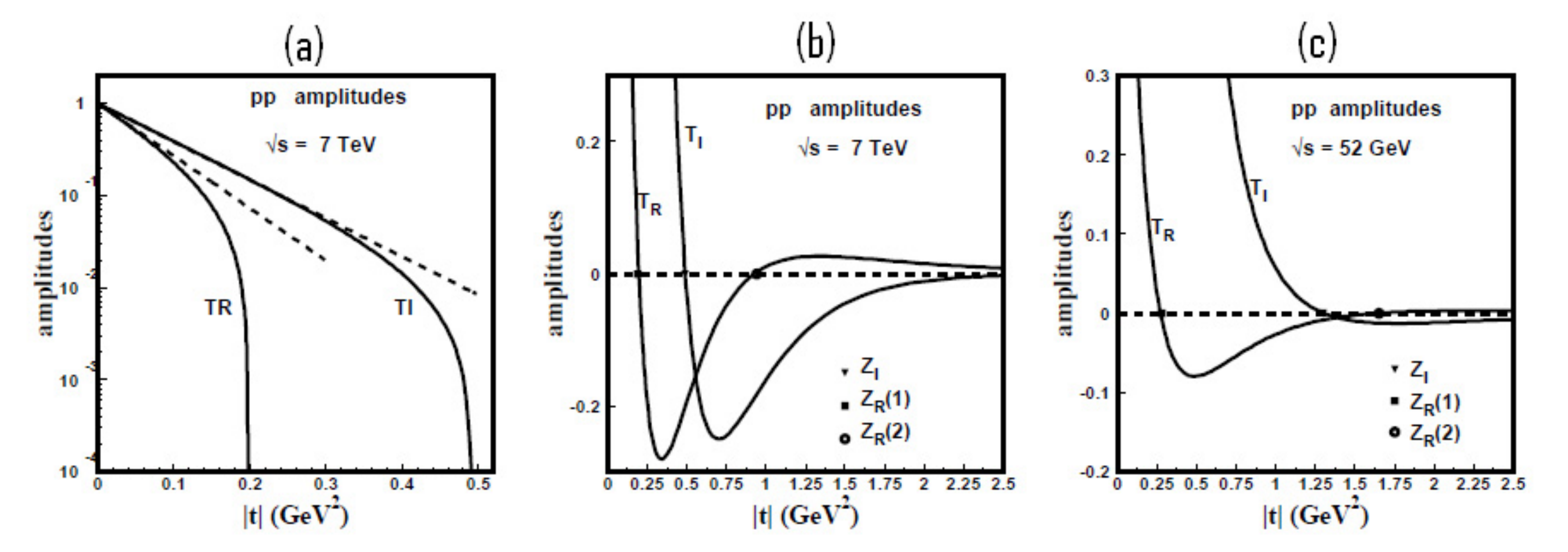}
\caption{ (a) Forward scattering
amplitudes $T_{R}$ and $T_{I}$ at $\sqrt{s} = 7 $ TeV in log scale, normalized
to one at $|t|=0$, showing their slopes, $B_{I}=19.77 $ and $B_{R}=30.2
\nobreak\,\mbox{GeV}^{-2}$, and their curvatures, and indicating positions of
the first zeros $Z_{R}(1)$ and $Z_{I}$; 
(b) Large-t dependence of the real and
imaginary scattering amplitudes showing the complete set of zeros; 
(c) t dependence of the real and imaginary scattering amplitudes at
$\sqrt{s}= 52.8$ GeV. Comparing with the figure for 7 TeV, we observe that 
all zeros  move  towards smaller  $|t|$ values as the energy increases. } 
\label{zeros} 
\end{figure*}

The amplitudes obtained in the analysis of the data, based on Eqs.
(\ref{ampTK}, \ref{cd1_s}), are shown in Fig. \ref{zeros} and numerical
information is given in Tables \ref{tableone} and \ref{tabletwo}. Their
general features are common to the lower energies from ISR and the Tevatron,
with regular variation of the parameters. The results agree with requirements
from dispersion relations for amplitudes and for slopes \cite{ferreira2}, and
the Coulomb interference accounts for the necessary generalization of the
Coulomb phase, presented in the Appendix. Near $t\simeq0$, the real part obeys
the theorem by A. Martin \cite{Martin} about its first zero, decreasing
quickly and crossing zero at small $|t|$ , before the imaginary part becomes
small. The $|t|$ dependence of the amplitudes for all $|t|$ is shown in part
(b) of Fig. \ref{zeros}. 
The imaginary part starts dominant over the real part, crosses
zero at higher $|t|$, then remains negative and asymptotically tends to zero
from the negative side, while the real part crosses zero again near $|t|=0.8
~\nobreak\,\mbox{GeV}^{2}$ , becoming positive. After the second real zero,
namely for $|t|$ larger than about 1.0 $\nobreak\,\mbox{GeV}^{2}$ , the real
amplitude stays positive, without further oscillation, and for $|t| \geq2~
\nobreak\,\mbox{GeV}^{2}$ becomes increasingly dominant over the imaginary
part. Important qualitative feature of our description of the data is that for
large $|t|$ the magnitude of the imaginary part is smaller than the positive
real part.

This behavior is a regular continuation of the results obtained for the ISR
energies, and as an example we show in part (c) of Fig. \ref{zeros}
the amplitudes for 52.8 GeV. At this and higher energies \cite{ferreira1}
the $Z_{I}$ zero occurs between $Z_{R}(1)$ and $Z_{R}(2)$. The magnitude 
of $T_{I}(s,t)$ is in general dominant over the real part between $Z_{R}(1)$ and $Z_{R}(2)$, so that the dip is located near $Z_{I}$. As the energy 
varies, while the first zero of $T_{R}(s,t)$ stays almost constant in 
the $|t|$ range from 0.15 to 0.3 GeV$^{2}$, the positions of both 
$Z_{I}$ and $Z_{R}(2)$ move to the left. The relative proximity of real and imaginary zeros influences the shape and depth
of the dip. At 540 and 1800 GeV the imaginary zero is distant from both real
zeros, so that no dip is formed, and only an inflection is observed in 
the $d\sigma/dt$ curve. 
 
Table \ref{tabletwo} gives   the positions of the zeros of the amplitudes and
the characteristic observable quantities in pp elastic scattering at 7 TeV and
52.8 GeV.
\begin{table*}[ptb]
\caption{ Positions of zeros of the amplitudes, locations of the predicted dip
and bump, and ratio characterizing the shape of this structure. For comparison
that shows the regularity, we give the same information for $\sqrt{s} $ = 52.8
GeV. We observe that as the energy increases all zeros move towards smaller
$|t|$, with the imaginary zero moving faster, becoming more distant from the
the second real zero $Z_{R}(2)$ and closer to the first one $Z_{R}(1)$. }%
\label{tabletwo}
\begin{tabular*}
{\textwidth}[c]{@{\extracolsep{\fill}}ccccccccccccc}\hline
$\sqrt s$ & $\mathrm{{Z_{I}} }$ ~ & $\mathrm{Z_{R}(1)}$~ & $\mathrm{Z_{R}(2)
}$~ & $|t|_{\mathrm{dip}}$ & $|t|_{\mathrm{bp}}$ & $(d\sigma/dt)_{\mathrm{dip}%
} $ & $(d\sigma/dt)_{\mathrm{bp}}$ & ratio & $\sigma_{\mathrm{el}}$ &
$\sigma_{\mathrm{inel}} $ & $\sigma_{\mathrm{el}}/ \sigma$ & \\
GeV & GeV$^{2}$ & GeV$^{2}$ & GeV$^{2}$ & GeV$^{2}$ & GeV$^{2}$ & mb/GeV$^{2}$
& mb/GeV$^{2}$ & bp/dip & mb & mb &  & \\\hline
52.8 & 1.3083 & 0.2710 & 1.6157 & 1.3560 & 1.7947 & $1.9\times 10^{-5}$ &
$6.3\times 10^{-5}$ & 3.2805 & 7.4308 & 35.0591 & 0.1749 & \\
7000 & 0.4671 & 0.1641 & 0.8235 & 0.4847 & 0.6488 & 0.0153 & 0.0282 & 1.8383 &
25.54 & 73.11 & 0.26 & \\\hline
\end{tabular*}
\end{table*}

The details of the dip-bump structure carry information on the scattering
amplitudes, and are particularly sensitive to their relative behavior near
their zeros \cite{ferreira1}. Thus, a marked dip appears when one of the
amplitudes (real or imaginary) crosses the zero in the interval where the
other amplitude stays nearly constant with magnitude small compared to the
variation of the former. In a domain where one of the amplitudes is dominant,
the zero of the amplitude with smaller magnitude does not affect the
observed differential cross section, as happens in the region of the first
real zero ($\approx0.15 - 0.3 ~\nobreak\,\mbox{GeV}^{2} $).

\begin{figure}[b]
\caption{ Partial cross sections $d\sigma^{I}/dt$ and $d\sigma^{R}/dt$ as
functions of $|t|$ as calculated with the analytic forms of Eqs. (\ref{ampTK})
and (\ref{cd1_s}). The dip in the sum $d\sigma/dt$ (at 0.485 GeV$^{2}$) is
close to the zero of the imaginary part (at 0.467 GeV$^{2}$). }%
\label{dsdt_re_im_fig}%
\includegraphics[height=10.2cm]{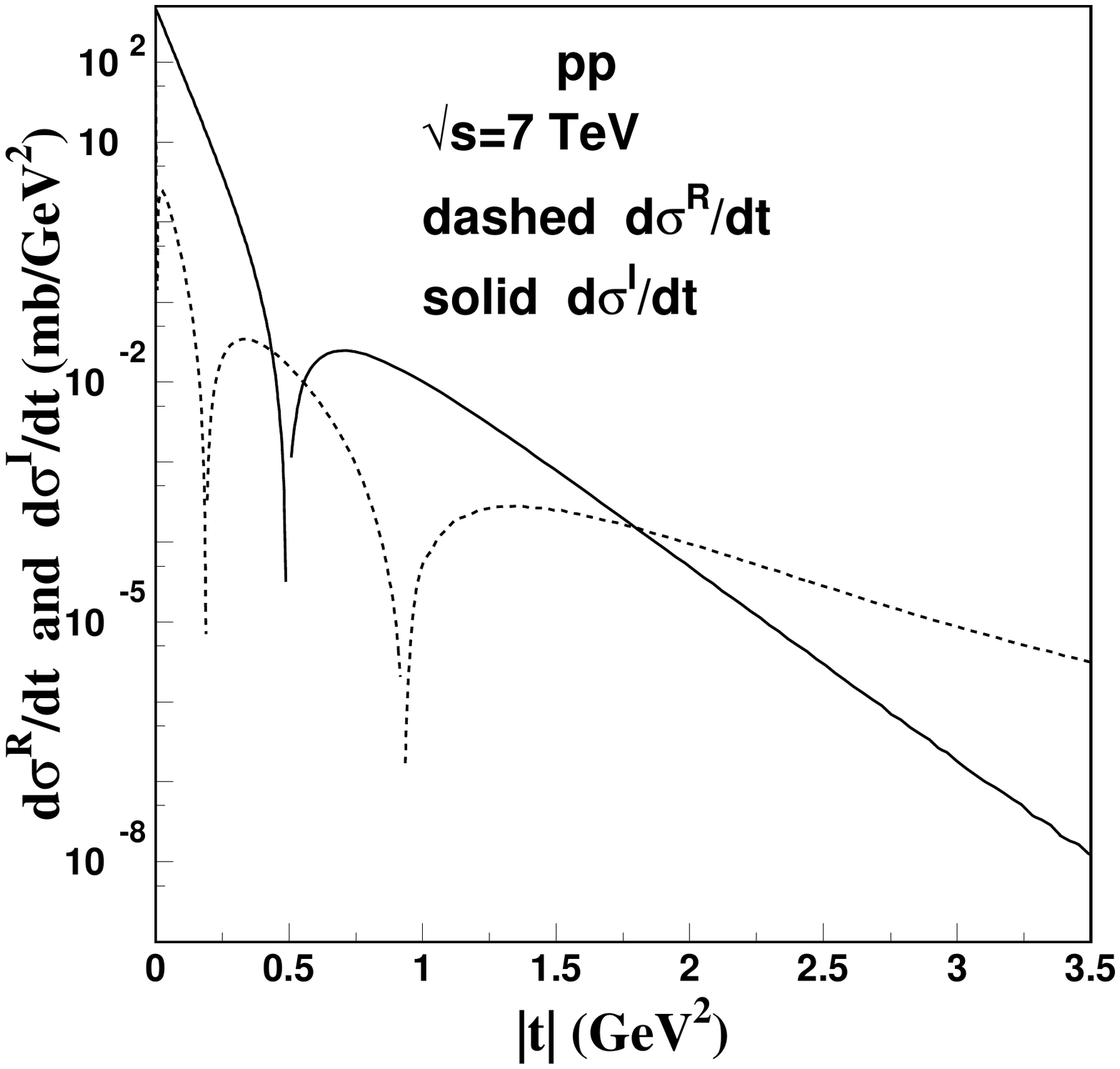}
\end{figure}

The parts $d\sigma^{R}/dt$ and $d\sigma^{I}/dt$ of the differential cross
section due to the real and imaginary amplitudes are shown in Fig.
\ref{dsdt_re_im_fig}. 

Differently from lower energies, the dip at $|t| \approx 0.5 ~
\nobreak\,\mbox{GeV}^{2} $ is more influenced by the proximity of the
imaginary zero ($ \approx 0.5 {\rm GeV}^{2}$) and the first real zero
$Z_{R}(1) $ (at $\approx$ 0.2 $\nobreak\, \mbox{GeV}^{2} $ ). Due to the increasing
proximity of the imaginary and real zeros at higher energies, the dip/bump
structure becomes more marked, and this is expected to happen at 14 TeV.

At 7 TeV the bump that follows the dip is formed by the $T_{I}$ and the $T_{R}
$ amplitudes that there have similar magnitudes, with $T_{I}$ becoming more
strongly negative while $T_{R}$ quickly becomes zero. This explains the ratio
nearly 2 (actually 1.838) between bump and dip heights. After the bump the
fall is faster because it is only determined by the dominant magnitude of the
imaginary amplitude, that decreases fast in $|t|$. $T_{R}$ is small positive,
but of long range, surviving until it meets the expected perturbative tail
after $|t|=2 ~ {\rm GeV}^{2}$.
\subsection{ Forward scattering}
The above description basically characterize our results of the
analysis of all TOTEM data  at 7 TeV, but the forward scattering part
deserves a  more  detailed  analysis, because the data in this
domain have higher precision and are crucial for the determination of the
total cross section.

To study separately the forward data investigating the use of amplitudes with
the exponential forms of Eq. (\ref{diffcross_eq}), we select a set of the
first 40 points in the interval $0.00515 \leq|t| \leq0.0907
\nobreak\,\mbox{GeV}^{2} $ . This forward set is used to compare the full-t
analytic form described above with forward scattering analyses using
exponential forms. Our analysis with Eqs. (\ref{ampTK}) and (\ref{cd1_s}), with
the parameters given in Eqs. (\ref{input}) and (\ref{other_param}), describes
the forward set of 40 points with an average squared deviation $\langle
\chi^{2}\rangle= 0.1062 $.

On the other hand, with Eq. (\ref{diffcross_eq}) and the forward scattering
parameters suggested by the TOTEM Collaboration\cite{talks}
\begin{eqnarray}
\sigma=98.58\pm2.23 ~\mathrm{mb} ,~\rho=0.141 \pm0.007~,\nonumber\\
~B_{I}=19.89\pm0.27~\mathrm{GeV}^{-2} , ~ B_{R}=B_{I} ~ ,\label{forward_exp}%
\end{eqnarray}
we find $\langle\chi^{2}\rangle= 0.1129 $ for this set of 40 points.

As a curiosity, we tested the pure exponential forms for the set of 40
points, using the values of $\sigma$ ,  $\rho$ , $B_{I}$ , 
 $B_{R}$  given in Eq. (\ref{input}), remarking  that they are
obtained in the all-t analysis. We then obtain an average $\langle\chi
^{2}\rangle=0.1034$, which may even appear as a better description
(actually, we can only say not worse) than obtained with the numbers of Eq.
(\ref{forward_exp}). 

\begin{figure}[ptb]
\caption{(a): Data and analytic representation for elastic pp scattering at 7
TeV in the forward range. Solid and dashed curves represent respectively the
large-t solution and the exponential forms of Eq. (\ref{diffcross_eq}) with
parameters of Eq. (\ref{forward_exp}) given by the experimentalists; (b) : The
same functions are drawn together with the 87 points of first part of TOTEM
data, showing the separation of exponential and full-t forms as $|t|$ becomes
large.}%
\label{forward_fig}%
\includegraphics[height=15.0cm]{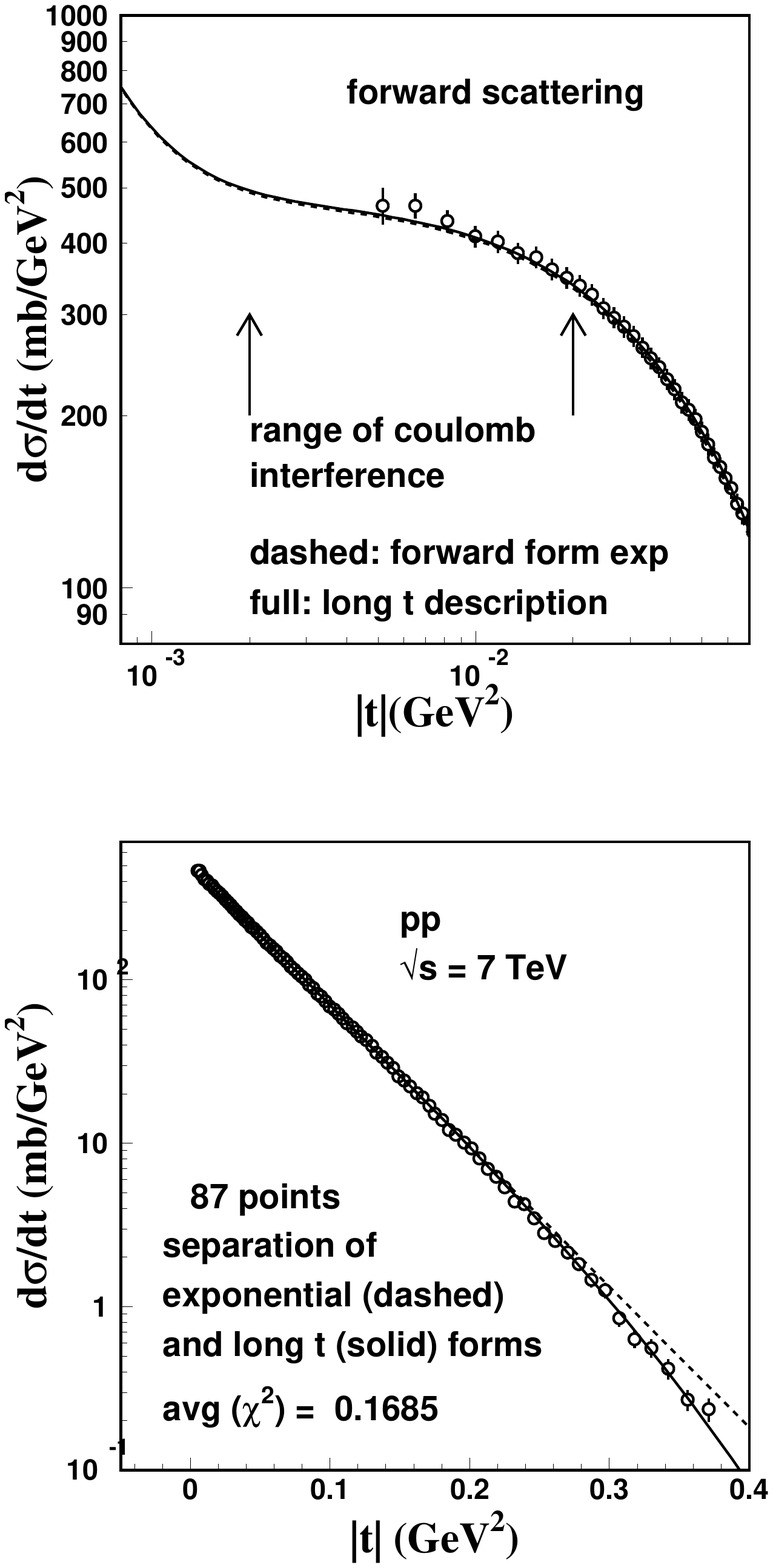}
\end{figure}

 Of course there is no statistical significance in these differences of
$\langle\chi^{2}\rangle$  values, and visually these representations
are almost degenerated as shown in part (a) of Fig. \ref{forward_fig}. The
importance of the construction of the full-t analytical forms is that they are
physically more realistic, including curvatures, magnitudes, signs, zeros of
the two amplitudes, reflecting important information on the collision dynamics.
In particular, one should remind that the usual treatment of data with  
$B_{R}=B_{I}$  is essentially wrong, although it is used in practice
because  $\rho$  is small. In fact, the slope  $B$  measured
directly in $d\sigma/dt$  is related to the slopes of the amplitudes
through 
\begin{equation}
B=\frac{B_{I}+\rho^{2}B_{R}}{1+\rho^{2}}~.\label{global_slope}%
\end{equation}
 Therefore, the expected difference between values is approximately
 $B_{I}\approx B-0.1\nobreak\,\mbox{GeV}^{-2}$ . However, such a
simplification hides important properties of the dynamics.  Furthermore, since
$d\sigma/dt$  is not a pure exponential form $A~e^{-B|t|}$,
determinations of $\sigma$  made with this approximation carry an
intrinsic error. The scattering amplitudes must have zeros, consequently the
slopes of the real and imaginary amplitudes are  $t$  dependent,
deviating from pure exponential forms. So, the determination with high
precision of total cross-section,  $\rho$  and slopes cannot be free from 
model-dependence, and differences are sensitive unless the measurements reach
very small  $|t|$ . Such a limitation of the approximation is shown in
part(b) of Fig. \ref{forward_fig}, where we exhibit the large-t and the
exponential solutions in the presence of the 87 points of the first part of
TOTEM measurements, with  $|t|\leq0.371~\nobreak\,\mbox{GeV}^{2}$. A
pure exponential form (dashed line) appears as adequate (visually,
and within errors) within the range  $\left\vert t\right\vert <0.25$,
but in reality the real part has a zero and becomes negative. 
   We here stress  that the physical quantities  $\sigma$ , $\rho$  , 
$B_{I}$  and $B_{R}$   are obtained as limiting values as 
    $|t|\rightarrow0$  from a complex
amplitude, thus requiring   a basic analytical form. Therefore, they
are naturally model-dependent, particularly if the data do no get close to the
origin. It is thus very meaningful that the numerical values in Eqs.
(\ref{input}) and (\ref{forward_exp}) are compatible. Note that our analytic
representation for the amplitudes covers the whole $t$  region with
the same accuracy, and contains more physical information than a mere
exponential fit. Naturally, the distinction between the two proposed
descriptions for $d\sigma/dt$, and consequently for the values of
$\sigma$ (98.65 and 98.58 mb) would require high precision in the
measurements. For example, for $|t|=5\times10^{-3}~{\rm GeV}^{2}$  the
predicted values for  $d\sigma/dt$  are respectively 
447.2 and 444.8   mb/GeV$^{2} $. 

 \subsection{Universality of the perturbative three gluon amplitude at 7 TeV}

At energies above $\sqrt{s} \approx20 ~ \nobreak\,\mbox{GeV} $, a perturbative
term due to three-gluon exchange, energy independent and of magnitude
$10^{-7}$ to $10^{-11}$ mb/GeV$^{2}$, produces a tail of form $|t|^{-8}$ that
dominates the scattering cross sections at large $|t|$. This term that was
observed in Faissler experiment at 27.4 GeV \cite{Faissler}, is real positive
for pp and negative for $\mathrm{{{p\bar{p}}}}$. In pp scattering at 52.8 GeV,
where data exist up to $|t|\approx10$ GeV$^{2}$, the matching of the data and
the tail is perfect and shows that the real amplitude at intermediate $|t|$
values, namely after the dip in the cross section, should actually be
positive, in agreement with our description of the amplitudes.

In our approach, the non-perturbative contributions vanish fast for large $|t|$, 
and we expect that  for $|t|\gg3~\mathrm{GeV}^{2}$, the cross sections at all
energies   behave like
\begin{equation}
d\sigma/dt \approx\left(  \hbar c\right)  ^{2} ~[R_{ggg} (t)]^{2}
\approx0.08~t^{-8}\ \ (\text{mb/GeV}^{2})~.\label{Asymptotic3g}%
\end{equation}

\begin{figure}[ptb]
\caption{(a) Prediction for the observation of the large $|t|$ tail
(triangles) at 7 TeV, based on the comparison with the data at 52.8 GeV. The
solid line is our solution for 7 TeV data with the tail added in the real part
amplitude. In part (b) the dashed line is our solution without the
perturbative tail. }%
\label{tail}%
\includegraphics[height=15.0cm]{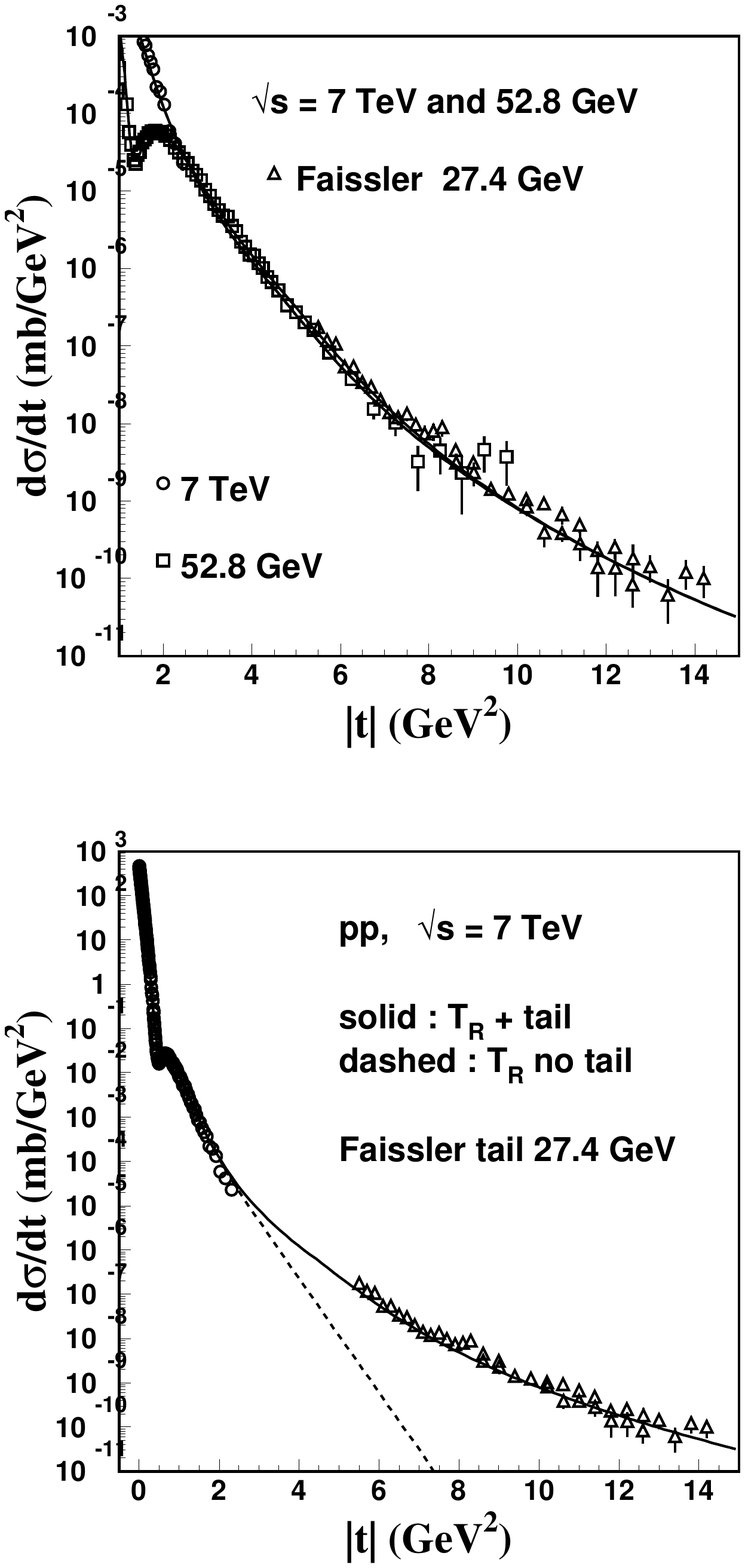}
\end{figure}

Fig. \ref{tail} shows part of the 52.8 GeV and 7 TeV data, together with the
points of the 27.4 GeV tail at large $|t|$. The superposition of points observed
starting at $|t|=2 ~ \nobreak\,\mbox{GeV}^{2}$ leads to the feeling that with
extended data the universal tail would be observed at 7 TeV, similarly to the
behavior seen at lower energies. The second part of the figure   shows 
the expected smooth connection of the TOTEM data at
7 TeV to the 27.4 GeV tail data points, according to the construction of the
amplitudes described above. According to this construction based on the assumption
of universality, the $|t|$ distribution is expected to reach the
three-gluon-exchange tail in the range from 5 to 15 GeV$^{2}$. Unfortunately
the experimental data end a little before, but observing the figure we are lead 
to believe that the present data are pointing towards the tail. Since $T_{R} (t)$
and the tail are positive, there will be no marked structure caused by
cancellation, as expected to occur in $\mathrm{p \bar p }$. At large $|t|$ the
cross section will be of form $1/|t|^{8}$ , according to the perturbative
three-gluon exchange tail. The superposition between $T_{R}$ an the
perturbative tail may provide precious information about the non-perturbative
regime. The real tail, with known sign, magnitude and shape, provides a
reference basis, analogous to what the Coulomb interaction does at small $|t|$, 
important for the identification of the amplitudes.

\subsection{ Impact Parameter Form for the Amplitudes   
\label{eikonal_form} }

 As mentioned before, our amplitudes given as functions of the
momentum transfer $|t|$ by Eqs. (\ref{ampTK}, \ref{cd1_s}) are written in
impact parameter space through the Fourier transforms
\[
\tilde{T}_{K}(s,b)=\frac{1}{2\pi}\int d^{2}\vec{q}~e^{-i\vec{q}.\vec{b}}%
~T_{K}^{N}(s,t=-q^{2})~
\]
where $K=I$ and $K=R$ in $\tilde{T}_{K}(s,b)$ refer to the real and imaginary
parts . They are given in analytical forms by
\[
\tilde{T}_{K}(s,b)=\frac{\alpha_{K}}{2~\beta_{K}}e^{-\frac{b^{2}}{4\beta_{K}}%
}+\lambda_{K}~\tilde{\psi}_{K}(s,b)~,
\]
where
\[
\tilde{\psi}_{K}(s,b)=\frac{2~e^{\gamma_{K}}}{a_{0}}~\frac{e^{-\sqrt
{\gamma_{K}^{2}+\frac{b^{2}}{a_{0}}}}}{\sqrt{\gamma_{K}^{2}+\frac{b^{2}}%
{a_{0}}}}\Big[1-e^{\gamma_{K}}~e^{-\sqrt{\gamma_{K}^{2}+\frac{b^{2}}{a_{0}}}%
}\Big]~
\]
is the shape function in $b$ space. Use has been made of the integration
formula \cite{ferreira1, Iwanami}
\begin{equation}
\int_{0}^{\infty}J_{0}(\beta v)~\frac{e^{-\lambda\sqrt{1+v^{2}}}}%
{\sqrt{1+v^{2}}}~v~dv~=~\frac{e^{-\sqrt{\lambda^{2}+\beta^{2}}}}{\sqrt
{\lambda^{2}+\beta^{2}}}~.
\end{equation}

\begin{figure}[b]
\caption{ (a) and (b) : respectively real and imaginary parts of the amplitude
in impact parameter space , showing separately the contributions from the
exponential ( Gaussian form, solid) and shape functions (dashed) of the
original amplitudes, and their sum (dotted). }%
\label{eikonals_ampl}%
\includegraphics[width=10.2cm]{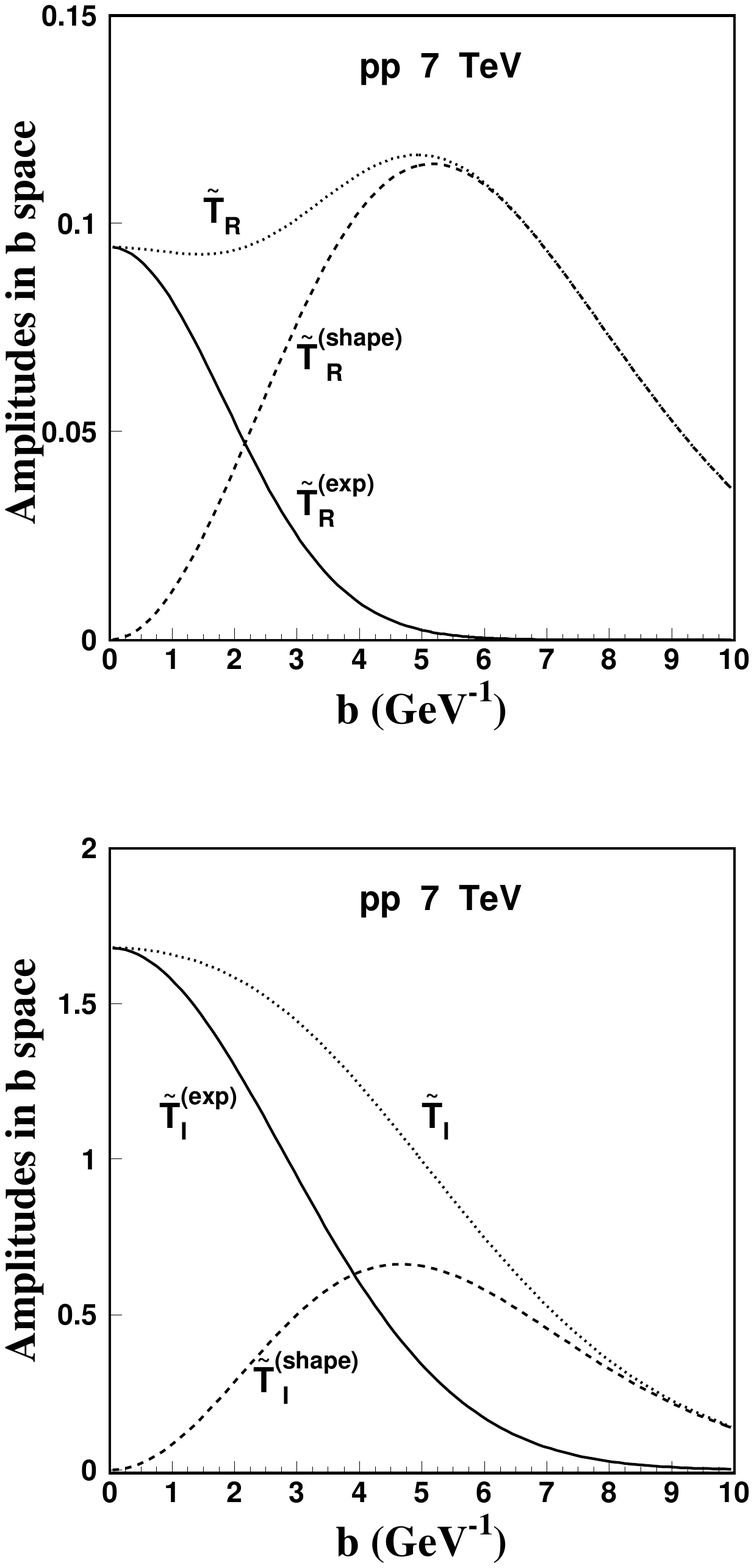}
\end{figure}

\begin{figure}[b]
\caption{Real and imaginary parts of the amplitude in impact parameter
coordinate.}%
\label{eikonal_sum}%
\includegraphics[height=10.2cm]{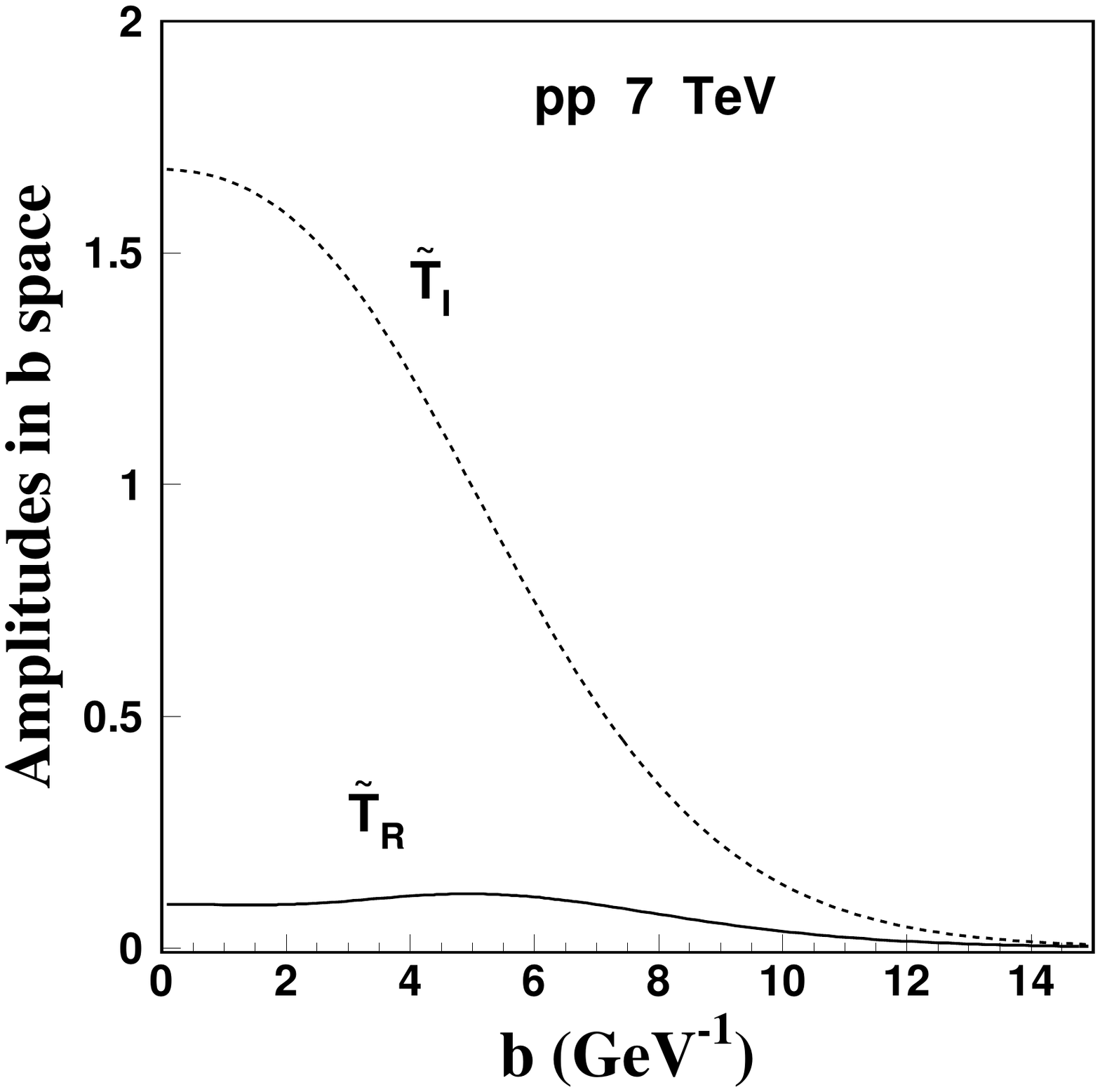}\end{figure}

Fig. \ref{eikonals_ampl} shows the parts $\tilde T_{K}^{\mathrm{(exp)}}(s,b)$
and $\tilde T_{K}^{\mathrm{(shape)}}(s,b)$ corresponding respectively to the
terms in exponential and shape function of the amplitudes in $|t|$ , and Fig.
\ref{eikonal_sum} shows their sums. 
 The shape
functions are zero at $b=0$ and all parts are positive for all $b$. The
amplitude in $b$-space can be put in the form
\begin{eqnarray}
\tilde T_{R}(s,b)~ + ~ i \tilde T_{I}(s,b) = i ~\sqrt{\pi} ~[ 1-
\mathrm{{e}^{2i\delta(s,b)}]}%
\end{eqnarray}
with real $\delta(s,b)$.

\begin{figure}[b]
\caption{Representation of the amplitude for the perturbative tail in $b$
space.}%
\label{eik_tail}%
\includegraphics[height=10.2cm]{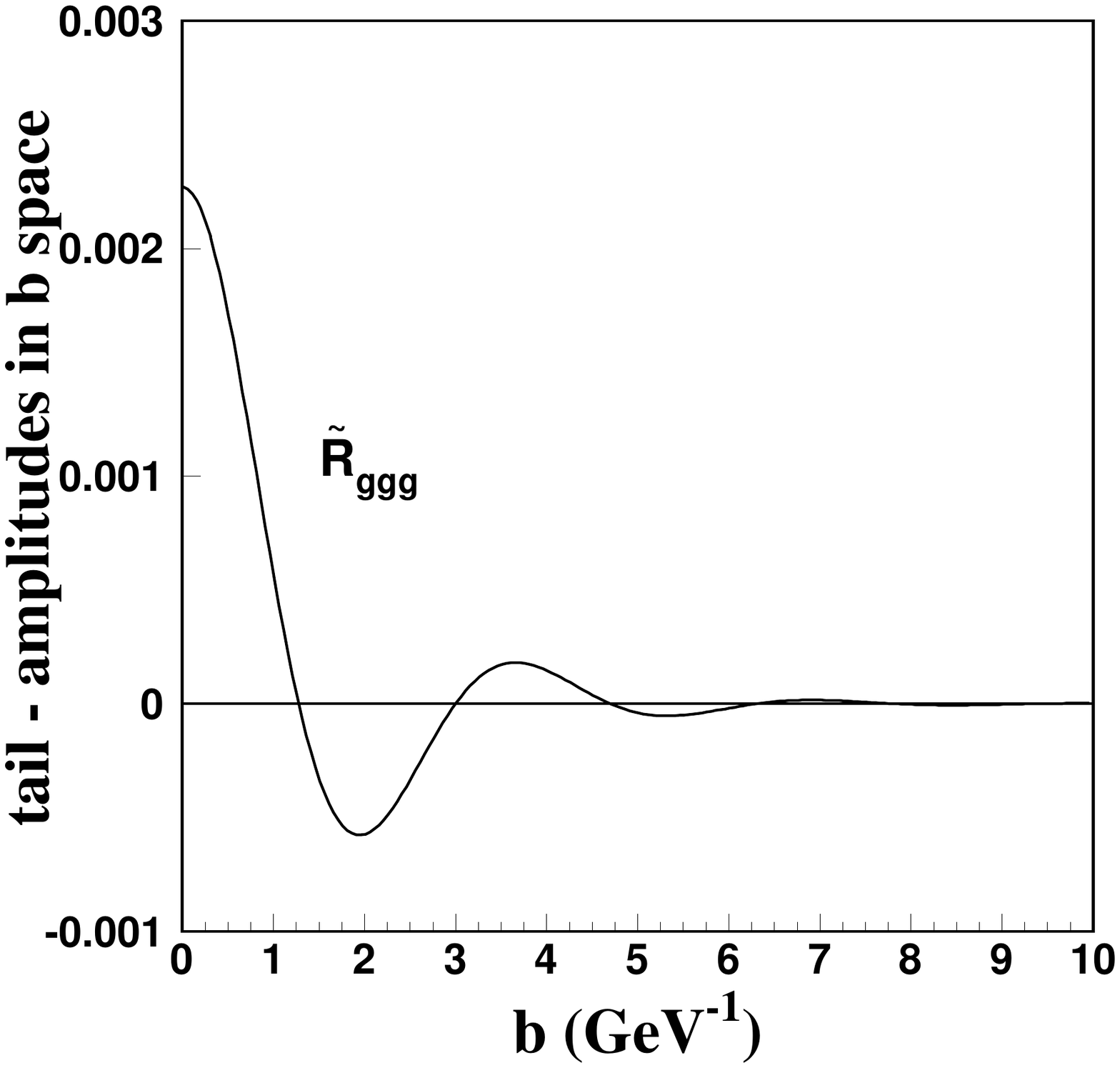}\end{figure}

The Fourier transform $\tilde R_{ggg}(s,b)$ of the perturbative tail is shown
if Fig. \ref{eik_tail}. Its behavior is of a form approximated like $\tilde
R_{ggg}(s,b) \approx0.0022 ~ J_{0}(c~b)/(1+(d~b)^{2})$ , with $c~ \approx~
1.863 $ and $d ~ \approx~4.0 $ in GeV. 

\subsection{Comparison with the amplitudes in the BSW model \label{BSW_model}}

Since the behavior of the amplitudes gives physical information on the collision
dynamics and its determination is model dependent, it is
important to compare our results with   other models
  that describe elastic pp and $p\bar{{p}}$ scattering processes 
\cite{BSW,Petrov1,Menon,Islam,Jenkovsky}. 
Results on amplitudes that  can be directly compared with ours
are given by the model proposed by Bourrely, Soffer and Wu (hereafter referred
to as BSW model) \cite{BSW}.  We find important
similarities and differences which we discuss below.

\begin{figure}[ptb]
\includegraphics[height=10.2cm]{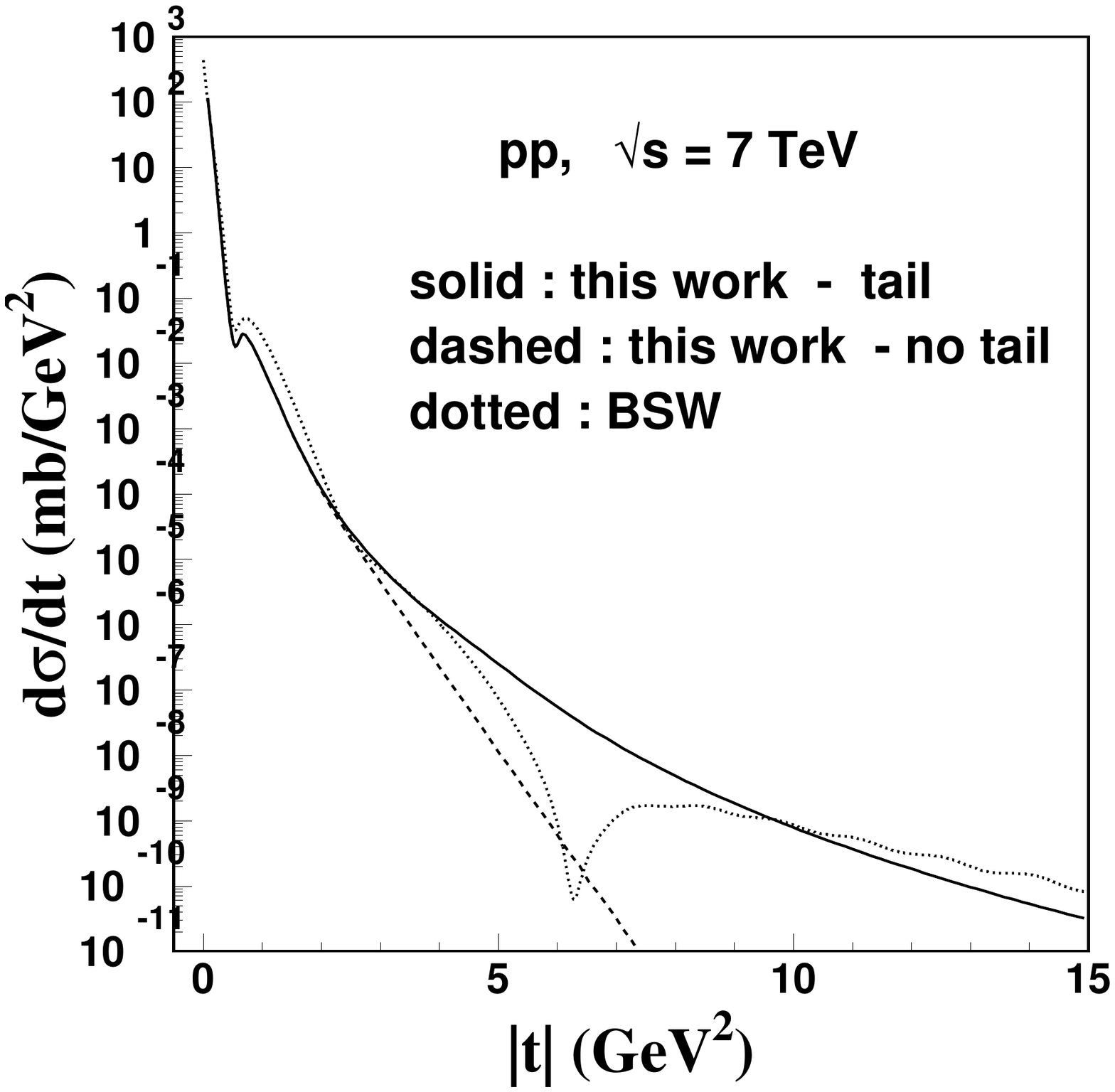}
\includegraphics[height=7.0cm]{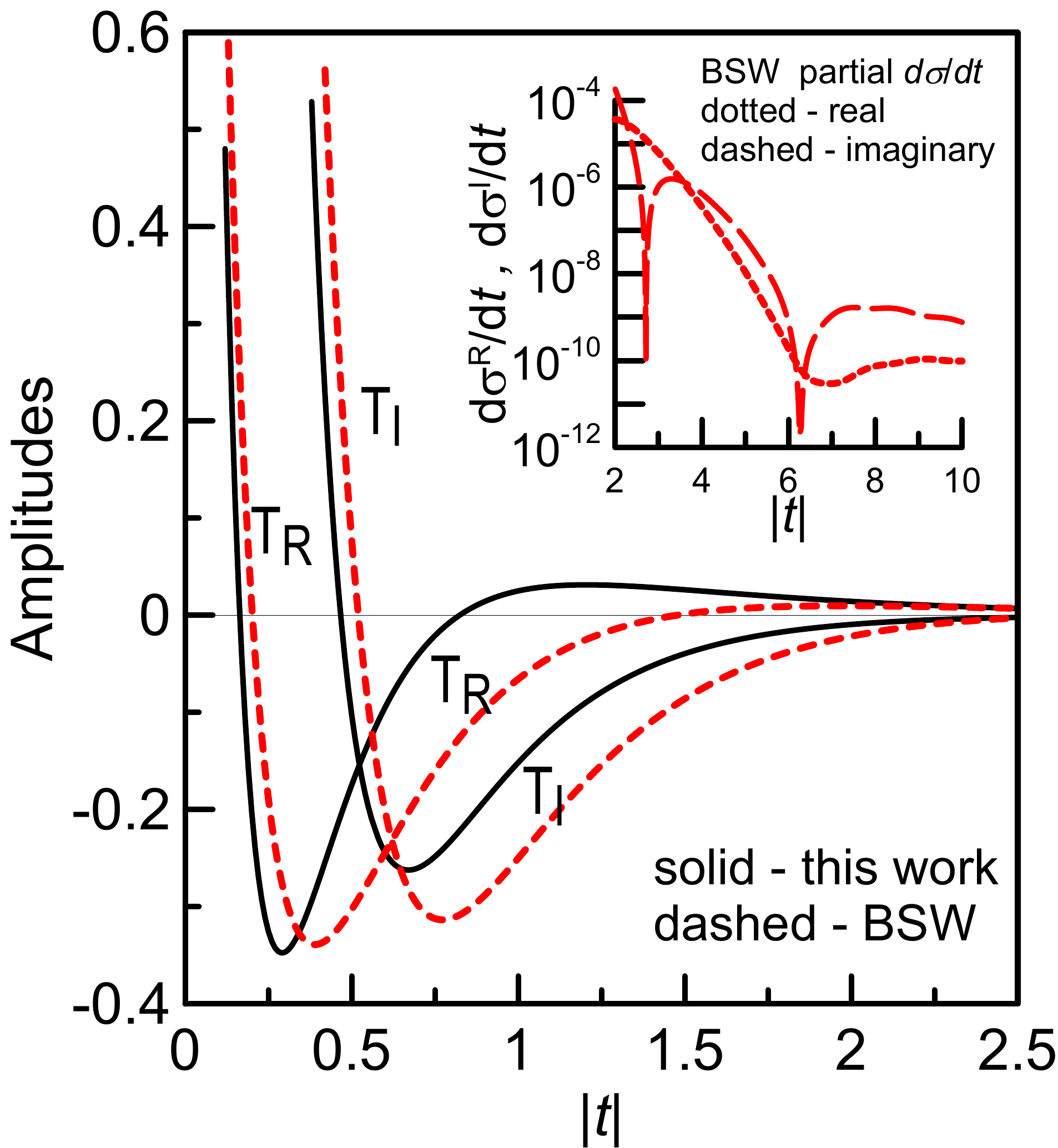}
\caption{ (a)
Comparison of $d\sigma/dt$ at 7 TeV in our description (solid line) and in BSW
model \cite{BSW} (dotted line). In the
large $|t|$ range our non-perturbative amplitudes fall exponentially (dashed
line) and the universal (for all energies) real three-gluon tail enters to
dominate the scattering. In the BSW case a second imaginary zero 
at $|t|\approx 2.7~\mathrm{GeV}^{2}$ causes no mark, because the real part
is not small, and a third zero in the imaginary amplitude predicts a marked 
dip  at $|t|\approx 6.3~\mathrm{GeV}^{2}$. 
(b) Comparison of amplitudes.
Solid lines show the real and imaginary amplitudes of our
representation, while dashed lines are for BSW.
 In the inset at up-right corner, we
show the large $\left\vert t\right\vert $ behavior of the amplitudes 
of the BSW model in terms of partial differential cross sections; 
two imaginary zeros occur  in this range. This can be compared to 
Fig. \ref{dsdt_re_im_fig}.}  
\label{BSW_7TeV}%
\end{figure}

The BSW model leads to values for the imaginary and real slopes with
$B_{R}>B_{I}$ (as expected), to a first real zero at small $|t|$ (as expected)
and to a first imaginary zero near the cross section dip (as expected) and to a second real zero at intermediate $|t|$. These
are the first basic, but crucial qualitative agreements with our approach. 
There also
exist several quantitative and qualitative differences that may reflect
important physical information behind the observables. In Fig. \ref{BSW_7TeV}
we compare details of cross section (part (a)) and of amplitudes (part(b)).

In the very forward region, the two descriptions cross each other, leading to
different values at the optical point (with total cross section 92.33 mb in
BSW and 98.65 mb for us) ; the slopes are qualitatively similar, with stronger
$B_{R}$ leading to a first real zero at low $|t|$ in both cases, as expected
by Martin's theorem.

In the intermediate domain ($0.3<\left\vert t\right\vert \,<2$) that includes
the dip-bump region, the BSW cross section overshoots the experimental data
and our results by a factor of nearly 2, caused
by excessive magnitudes of both real and imaginary
amplitudes in this region.
The location of the dip-bump structure is also slightly shifted to a larger
$\left\vert t\right\vert $ and the BSW dip is broader than  ours. For
$\left\vert t\right\vert \simeq2.5~ \mathrm{GeV}^{2},$ the two representations 
approach each other, but this is rather a coincidence as we can see from
the behavior of the amplitudes.

After the first zeros, the BSW amplitudes have larger magnitudes, and the
second real zero is shifted to larger $\left\vert t\right\vert $ compared to
ours, and this has influence in the dip-bump shape.
Near the first zero of the imaginary amplitude, the BSW real amplitude is
about 70\% larger than ours in magnitude.

 Characteristic quantities of the BSW
results are  $~\sigma=$ 92.33 mb, $B_{I}=19.3~\mathrm{GeV}^{-2}$,
$B_{R}=25.8~\mathrm{GeV}^{-2}$, $\rho=0.126$, to be compared to the values
given in Table \ref{tableone}. 
In the range of interest their amplitudes have 
two real zeros, at 0.205 and 1.475 (a third real zero is farther away), and
three imaginary zeros at 0.525, 2.705 and 6.265 ; in our case there are two
(and only two) real zeros, both more to the to the left, and only one
imaginary zero, also at smaller $|t|$.

Although the quantitative differences are important, it is impressive the
similarity in the forms of the amplitudes in the two calculations in the 
data range, up to 2.5 GeV$^{2}$. This is the region  where we believe that the
non-perturbative dynamics is dominant and where eikonal-based models are
expected to be valid.

Important qualitative differences appear in the
behavior of amplitudes for large $|t|$, namely   $
|t| \gtrsim 2 \rm{GeV}^{2}  $, which is illustrated
in the form of partial contributions to the cross section of the BSW model 
in the up-right corner inset of Fig.\ref{BSW_7TeV}-(b).  
 While in our model (solid for real and
dash-dotted for imaginary), the non-perturbative amplitudes fall off
exponentially for large $|t|$ and the the real part is
made dominant with the action of the perturbative three-gluon exchange term,
in BSW the amplitudes fall rather slowly, and exibit further zeros.
Contrary to our approach, BSW cross section is dominated by the imaginary part
for large $|t|.$ Due to the presence of the second zero,
there appears a slight inflection at $|t| \sim3~{\rm GeV}^{2}.$
Interestingly such slight inflection also appears in our model coincidentally 
by a completely different reason, due to the perturbative real amplitude
summing to the real amplitude.  It seems that $\sqrt{s}=$ 52 GeV data supports
strongly the contribution of this three-gluon exchange amplitude
that we include in our description.

\begin{figure}[ptb]
\includegraphics[height=10.2 cm]{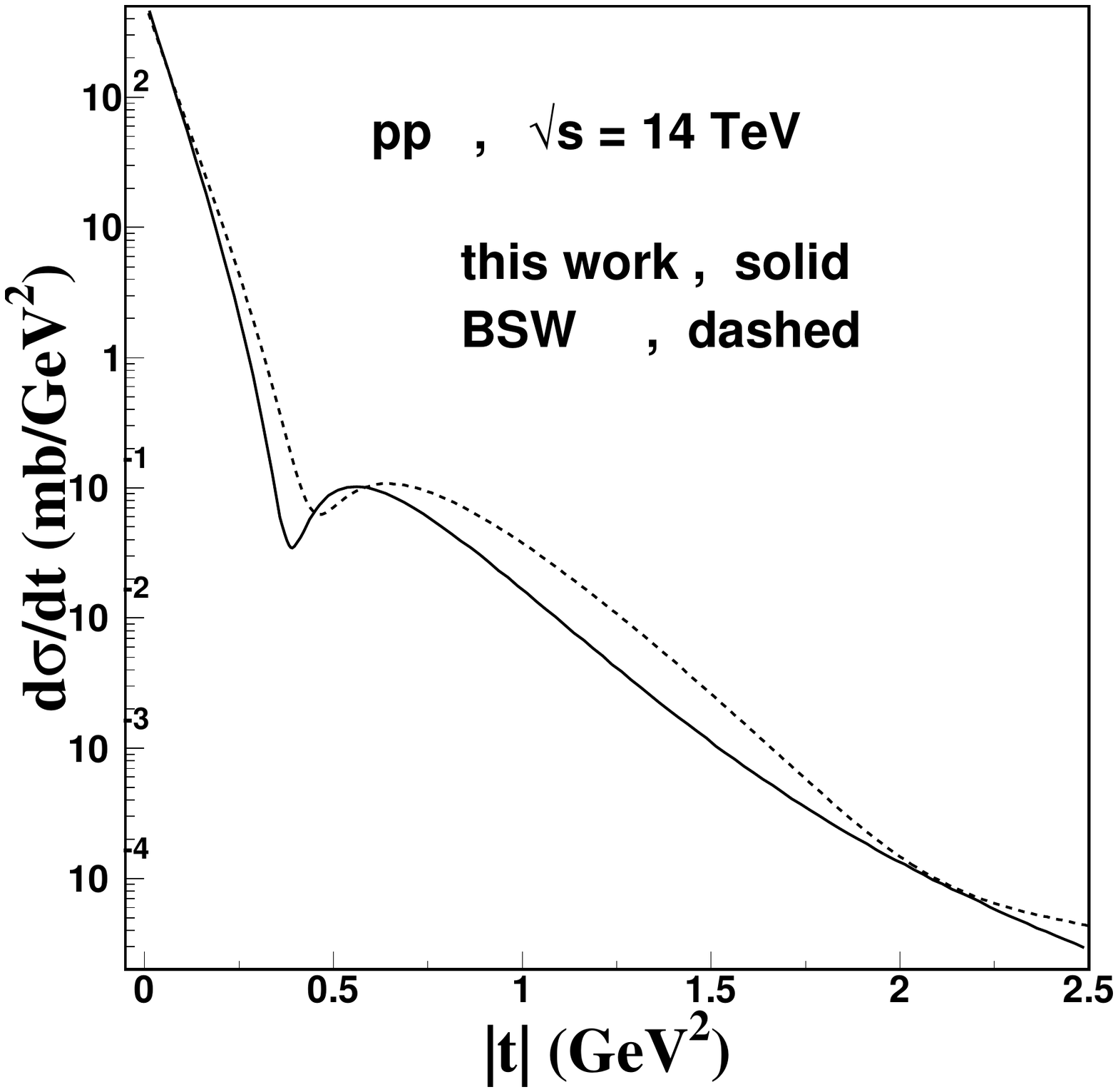}\caption{ Comparison of
predictions for 14 TeV , where the dip/bump structure becomes more pronounced.
}%
\label{prediction_14}%
\end{figure}

The above described distinctions between BSW and our model lead to significant
differences in the expectations for 14 TeV. In our case, we use the smoothness
in the energy dependence of the parameters and of the characteristic features
of the amplitudes to obtain a prediction for the higher energy, that shows a
pronounced dip-bump structure, which  appears at a bit smaller
$\left\vert t\right\vert .$ This is due to the zero of the imaginary part,
pinched by the two zeros of real part which are both approaching the imaginary
zero for large $\sqrt{s}$. In Fig.\ref{prediction_14}, we compare the two predictions, which are quite
different in the behavior of dip-bump structure, as well as in the large
$\left\vert t\right\vert $ behavior.


\section{Summary and Discussion \label{summary}}

In this paper we present an analysis of the recent TOTEM data for the elastic
scattering cross section at $\sqrt{s}=7$ TeV, extending method employed at
lower (ISR and Fermilab) energies \cite{ferreira1}. The central point of our
approach is that we describe the differential cross section directly in terms of
scattering amplitudes, expressed analytically for all observed $|t|$ domain,
together with the contribution from the perturbative three-gluon exchange
process at large $|t|$. Differently from a mere parameterization fit of the
differential cross section, our approach offers a qualified interpretation of
data in terms of amplitudes, intending to serve as a bridge between data and
theoretical descriptions by dynamical models.

For example, it is fundamental to recognize the difference of the slope
parameters of real and imaginary amplitudes to describe the all $|t|$ domain,
although, due to the small value of the $\rho$ parameter the effect is not clearly visible while looking only at the forward
scattering data. Furthermore, our analytic representation can also be written explicitly
in the impact parameter space, that may permit us to discuss the physical
meaning of our parameters in terms of geometrical models.

Our approach works perfectly well for $\sqrt{s}=7$ TeV , giving an excellent
representation for the data, including the forward region, and covering the
dip at $|t|\approx0.5 ~ \nobreak\,\mbox{GeV}^{2}$ and the subsequent bump. The
analysis leads to a description of the dip/bump structure of the differential
cross section in terms of positions of zeros and magnitudes of the real and
imaginary amplitudes.

At 7 TeV the zero of $T_{I}$ and the dip nearly coincide, at about 0.5
GeV$^{2}$, so that the height of the dip informs  the magnitude of $T_{R}$ in
this region. In the bump region after the dip, the magnitudes of the real and
imaginary parts become similar, so that the form of the subsequent bump
structure depends very sensitively on the interplay of the two amplitudes. The
maximum of the bump occurs where the magnitudes of the real and imaginary
amplitudes are nearly equal. With this behavior, the ratio bump/dip = 1.84 is
nearly 2 .

The gap in energy of the experiments at 7 TeV and the energies accessed in the
long history of pp and $\mathrm{p\bar{p}}$ scattering is enormous.
Nevertheless, elastic (soft) processes are very conservative, with smooth
energy dependences, so that the experience, methods and knowledge acquired 
before can be transferred  to this higher energy domain. 
The zeros of the amplitudes
determined from the present analysis (one imaginary zero and two real zeros)
are consistent as natural extension of the previous work for lower energies.
Since the behavior of the free parameters is smooth, we can extend the
present analysis to the future $\sqrt{s}=14$ TeV experiment. We foresee that
there will be a clear dip at $t\simeq0.45~$GeV$^{2}$ with similar bump structure with 
ratio $\approx 3$ and a total cross section $\sigma\simeq108$ mb.  This description
concerns the non-perturbative regime characteristic of elastic and diffractive
scattering (including multi-pomeron and reggeon exchanges).

It also becomes clear that the assumption, introduced and discussed by
Donnachie and Landshoff \cite{landshoff} on the universality of the three
gluon exchange amplitude, observed at 27.4 GeV \cite{Faissler} and higher ISR
energies, is still consistent with the 7 TeV data, that likely are pointing to
the distant tail, as seen in Fig. \ref{tail}. If this assumption is true,
some very interesting consequences can be drawn. First, for $\left\vert
t\right\vert \gg3~ \nobreak\,\mbox{GeV}^{2}$, the scattering amplitude is
dominated by this process and consequently the imaginary part should be small,
while the magnitude and sign of the real part is determined by Eq.
(\ref{R_tail}). Although the present analysis already indicates the
consistency of this assumption, the precise behavior of the amplitudes in the
transition region from non-perturbative to perturbative dynamics may be
affected sensitively by the data in this region. Therefore, it is extremely
important to confirm this behavior at $\sqrt{s}~=$ 7 TeV by extending TOTEM
analysis to higher $|t|$ values, with consequences for the identification of
the amplitudes in the range of transition. If $d\sigma/dt$ is found to follow
the tail without marked structure, we can conclude that in this region the
real amplitude has positive sign and that the imaginary part has much smaller
magnitude. It would be nice to reach more extended $|t|$ ranges and high
precision in this new era of studies of high energy collisions.\

It is interesting to remark that in the $p\bar{p}$ case, the negative 
perturbative term may produce a third
zero in the real amplitude causing a dip in the $|t|$ distribution.
This could have been seen in Fermilab data at 540 and 1800 GeV 
if the measurements had reached high enough $|t|$ values.  

In Subsec. \ref{BSW_model} we compare several aspects of our work 
with the results of a specific model  \cite{BSW}, showing   important 
similarities and differences, and stressing that comparison of models 
in the amplitude level is essential for the progress in the understanding 
of the scattering process.   

 To conclude, working with a specific analytical form for the
amplitudes, we have produced a detailed and precise description of the data of
elastic pp scattering at 7 TeV. The knowledge of the individual amplitudes
carries more physical information on the dynamics of the scattering processes,
and this work is part of an effort to find a consistent description of the
amplitudes covering regularly the data for all energies in the whole t - range
\cite{ferreira1}. The present analysis of the TOTEM data reproduces accurately
the behavior of the observed cross section in the whole $t$-range,
with consistent proposal for the determination of the amplitudes, confirming
the expectations of similarity with lower energies. It is hoped that this
description can be successfully extended to the future measurements at higher
energies. 

\section{Appendix: The Coulomb Phase \label{phase_App}}


Here we present  an expression for the Coulomb interference phase appropriate
for forward scattering amplitudes with $B_{R} \neq B_{I}$
\cite{KEK_2009}.

The starting point is the expression for the phase obtained by West and Yennie
\cite{WY}
\begin{equation}
\label{WYphase}\Phi(s,t)=(-/+)\Bigg[\ln\bigg(-\frac{t}{s}\bigg)+\int_{-4p^{2}%
}^{0} \frac{dt^{\prime}}{|t^{\prime}-t|}\bigg[1-\frac{F^{N}(s,t^{\prime}%
)}{F^{N}(s,t)}\bigg] \Bigg] ~ ,
\end{equation}
where the signs $(-/+)$ are applied to the choices pp/p$\bar{\mathrm{p}} $
respectively. The quantity $p$ is the proton momentum in center of mass
system, and at high energies $4p^{2} \approx s $ .

For small $|t|$, assuming that $F^{N}(s,t^{\prime})$ keeps the same form for
large $|t^{\prime}|$ (this approximation should not have practical importance
for the results), we have
\begin{eqnarray}
\frac{F^{N}(s,t^{\prime})}{F^{N}(s,t)}  & =\frac{F_{R}^{N}(s,0)e^{B_{R}%
t^{\prime}/2}+i~F_{I}^{N}(s,0)e^{B_{I}t^{\prime}/2}}{F_{R}^{N}(s,0)e^{B_{R}%
t/2}+i~F_{I}^{N}(s,0)e^{B_{I}t/2}}\nonumber\\
& =\frac{c}{c+i}~e^{B_{R}(t^{\prime}-t)/2}+\frac{i}{c+i}~e^{B_{I}(t^{\prime
}-t)/2}~,\label{relation1}%
\end{eqnarray}
where
\begin{equation}
c~\equiv~\rho e^{(B_{R}-B_{I})t/2}~.\label{def_c}%
\end{equation}
The integrals that appear in the evaluation of Eq. (\ref{WYphase}) are reduced
to the form \cite{KL}
\begin{equation}
I(B)=\int_{-4p^{2}}^{0}\frac{dt^{\prime}}{|t^{\prime}-t|}%
\bigg[1-e^{B(t^{\prime}-t)/2}\bigg]~,\label{int_form}%
\end{equation}
that is solved in terms of exponential integrals \cite{abramowitz} as
\begin{eqnarray}
I(B)= & E_{1}\big[\frac{B}{2}\bigg(4p^{2}+t\bigg)\big]-E_{i}\big[-\frac{Bt}%
{2}\big]+\ln\big[\frac{B}{2}\bigg(4p^{2}+t\bigg)\big]\nonumber\\
+  & \ln\big[-\frac{Bt}{2}\big]+2\gamma~.\label{functional_form}%
\end{eqnarray}

The real and imaginary parts of the phase are then written
\begin{equation}
\label{realphase}\Phi_{R}(s,t) =(-/+)\Bigg[\ln\bigg(-\frac{t}{s} \bigg) +
\frac{1}{c^{2}+1}\bigg[ c^{2} I(B_{R})+I(B_{I}) \bigg] \Bigg] ~ ,
\end{equation}
and
\begin{equation}
\label{imagphase}\Phi_{I}(s,t) =(-/+) \frac{c}{c^{2}+1} \Bigg[ I(B_{I}%
)-I(B_{R}) \Bigg] ~ .
\end{equation}

With $\sigma$ in mb and $t$ in GeV$^{2}$, the practical expression for
$d\sigma/dt$ in terms of the parameters $\sigma$, $\rho$ , $B_{I}$ and $B_{R}$
is
\begin{eqnarray}
\frac{d\sigma}{dt}=\pi\left(  \hbar c\right)  ^{2}~\Bigg[\bigg[\frac
{\rho~\sigma~e^{B_{R}t/2}}{4\pi\left(  \hbar c\right)  ^{2}}+F^{C}%
e^{\alpha\Phi_{I}}\cos(\alpha\Phi_{R})\bigg]^{2}\nonumber\\
+\bigg[\frac{\sigma~e^{B_{I}t/2}}{4\pi\left(  \hbar c\right)  ^{2}}%
+F^{C}e^{\alpha\Phi_{I}}\sin(\alpha\Phi_{R})\bigg]^{2}\Bigg]~.\label{dsigdt}%
\end{eqnarray}

At high energies and small $|t|$ we simplify $ 4p^{2}+t\rightarrow s $
and then the functional form of $I(B)$ is written
\begin{eqnarray}
I(B)=E_{1}\bigg(\frac{Bs}{2}\bigg)-E_{i}\bigg(-\frac{Bt}{2}\bigg)+\ln
\bigg(\frac{Bs}{2}\bigg)\nonumber\\
+\ln\bigg(-\frac{Bt}{2}\bigg)+2\gamma~.\label{functional_form2}%
\end{eqnarray}

For large $s$ , the term $E_{1}(Bs/2) $ can be neglected, and the phase
becomes insensitive to $s$.

The usual expression from West and Yennie
\begin{equation}
\alpha\Phi_{WY}=(-/+)\alpha\Bigg[\gamma+\ln\bigg(-\frac{Bt}{2}%
\bigg)\Bigg]~\label{WYphase-2}%
\end{equation}
can be obtained from Eq. (\ref{realphase})  with $B_R=B_I=B$ and 
 using the low $t$ behavior
  \[
E_{i}\big[-\frac{Bt}{2}\big] \approx\gamma+ \log-\frac{Bt}{2} ~.
\] 
We  recall that the value of the slope $B$ usually taken from the 
experimental $d\sigma/dt$ data is the average given by 
Eq. (\ref{global_slope}). 

We obtain \cite{KEK_2009}  that the values of  $\alpha\Phi_{I}(s,t)$ are very small 
 so that the imaginary part of the phase can be 
safely put equal do zero.

\begin{figure}[ptb]
\caption{Values of the Coulomb interference phase evaluated in the kinematical
conditions of pp elastic scattering at 7 TeV , with account made for the
difference in the slopes of the real and imaginary amplitudes. For comparison
we show the phase obtained with the basic formula \cite{Cahn} 
of Eq.(\ref{formula_cahn})}.
\label{fig_phase_7TeV}%
\includegraphics[height=10.2cm]{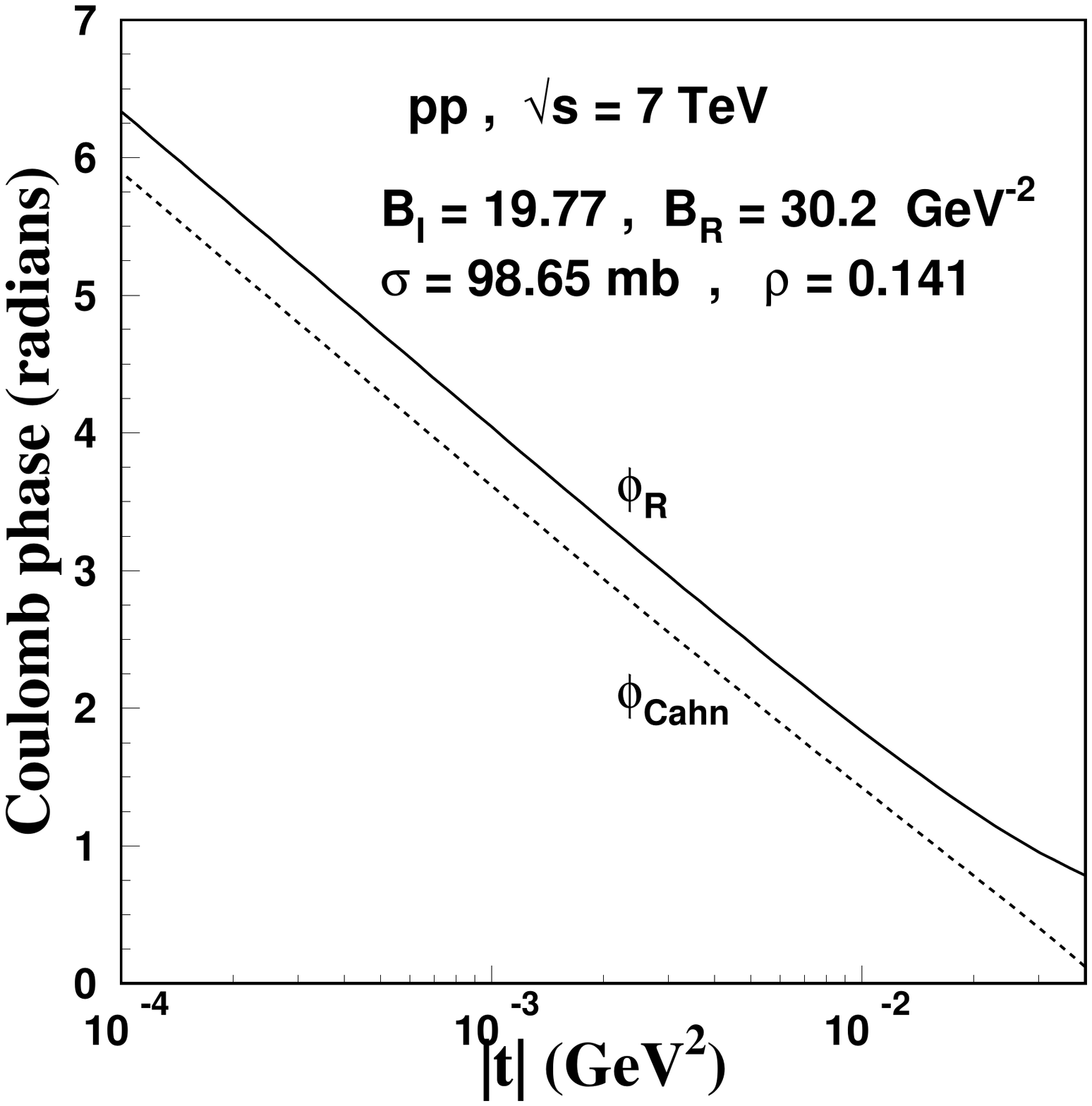}
\end{figure}

The construction of the Coulomb phase has been studied in the eikonal
formalism appropriate for the interference of Coulomb and nuclear interactions
\cite{KL,Cahn,Selyugin} with special attention given to the influence of
the proton electromagnetic form factor. These treatments keep the assumption
that the real part of the nuclear amplitude has the same slope as the
imaginary part, and the results are not very different from the West Yennie
formula. Some of these results have been tested against the data
\cite{Petrov2}.

Fig. \ref{fig_phase_7TeV} shows the values of the Coulomb interference phase
in the kinematical conditions of the 7 TeV LHC experiment. We show together
the phase obtained with the expression \cite{Cahn}
\begin{eqnarray}
\Phi_{\mathrm{Cahn}} = - \big[ \gamma+ \ln(-B t/2) + \ln\big(1 + 8/(B
\Lambda^{2})\big)   \\ \nonumber
+ (-4t/\Lambda^{2})\ln(-4t/\Lambda^{2}) -2t/\Lambda^{2} \big] ~, 
\label{formula_cahn}
\end{eqnarray}
where the proton form factor has been used with exponential form ,
$\Lambda^{2} = 0.71 \nobreak\,\mbox{GeV}^{2}$, and $B$ is taken as equal to
$B_{I}$.

\begin{acknowledgements}

The authors are very grateful to the members of the Totem Collaboration, particularly to 
K. Osterberg, S. Giani and M. Deile, for offering an opportunity of presentation and discussion of the present work.
The authors wish to thank CNPq, PRONEX and FAPERJ for financial support. A
part of this work has been done while TK stayed as a visiting professor at
EMMI-ExtreMe Matter Institute/GSI at FIAS, Johann Wolfgang Universit\"{a}t,
Frankfurt am Main. TK expresses his thanks to the hospitality of Profs. H.
Stoecker and D. Rischke.
\end{acknowledgements}

\end{document}